\documentclass[fleqn,10pt]{wlscirep}
\usepackage[utf8]{inputenc}
\usepackage[T1]{fontenc}
\usepackage{siunitx}
\usepackage{subcaption}
\usepackage{multirow}
\usepackage{multicol}
\usepackage{colortbl}
\usepackage{graphicx}
\usepackage{amssymb}
\usepackage[version=4]{mhchem}
\usepackage{hyperref}
\usepackage{xspace}

\newcommand{\nanogen}{\textsc{PXRDnet}\xspace}
\newcommand{\mppxrd}{\textsc{MP-20-PXRD}\xspace}
\newcommand{\rw}{$R_{wp}^{2}$\xspace}

\title{\textit{Ab Initio} Structure Solutions from Nanocrystalline Powder Diffraction Data via Diffusion Models}

\author[1*, 6]{Gabe Guo}
\author[1]{Tristan Luca Saidi}
\author[2]{Maxwell W. Terban}
\author[3]{Michele Valsecchi}
\author[4*]{Simon J. L. Billinge}
\author[5*]{Hod Lipson}
\affil[1]{Columbia University, Department of Computer Science, New York, NY, USA, 10027}
\affil[2]{Momentum Transfer, 68169 Mannheim, Germany}
\affil[3]{Columbia University, Department of Chemical Engineering, New York, NY, USA, 10027}
\affil[4]{Columbia University, Department of Applied Physics and Applied Mathematics, New York, NY, USA, 10027}
\affil[5]{Columbia University, Department of Mechanical Engineering, New York, NY, USA, 10027}
\affil[6]{Stanford University, Department of Computer Science, Stanford, CA, USA, 94305}

\affil[*]{Correspondence: \{gzg2104, sb2896, hod.lipson\}@columbia.edu, gabeguo@stanford.edu}

\begin{abstract}
A major challenge in materials science is the determination of the structure of nanometer sized objects. Here we present a novel approach that uses a generative machine learning model based on diffusion processes that is trained on 45,229 known structures. 
The model factors both the measured diffraction pattern as well as relevant statistical priors on the unit cell of atomic cluster structures. Conditioned only on the chemical formula and the information-scarce finite-size broadened powder diffraction pattern, we find that our model, \nanogen, can successfully solve simulated nanocrystals as small as 10~\AA\ across 200 materials of varying symmetry and complexity, including structures from all seven crystal systems. We show that our model can successfully and verifiably determine structural candidates four out of five times, with average error among these candidates being only 7\% (as measured by post-Rietveld refinement R-factor). Furthermore, \nanogen is capable of solving structures from noisy diffraction patterns gathered in real-world experiments. We suggest that data driven approaches, bootstrapped from theoretical simulation, will ultimately provide a path towards determining the structure of previously unsolved nano-materials. 

\end{abstract}
\begin{document}

\flushbottom
\maketitle
% * <john.hammersley@gmail.com> 2015-02-09T12:07:31.197Z:
%
%  Click the title above to edit the author information and abstract
%
%\thispagestyle{empty}

\section*{Introduction}

% Why do people care about this topic
The ongoing materials revolution of the past hundred years has been built upon the scientific community's heightened understanding of atomic arrangements (\textit{i.e.}, material structure) and the inherent dependence of material properties on this underlying structure. 
In regards to determining material structure, the \textit{conditio sine qua non} is single crystal structure solution. 
In this optimal setting,
despite the loss of phase information, the diffraction patterns contain maximal information content for the structure solution \cite{giacovazzo2002fundamentals} and are often successful. 
% Problem that prevents this topic from reaching full fruition
Yet, in increasingly many practical settings, it is not feasible to obtain a pure single-crystal sample. 
This is especially acute for nanometer-sized atomic clusters, the so-called nanostructure problem \cite{nano_billinge2007problem}.
In these cases the diffraction patterns have significant degradation in information content. Peak intensities must be extracted from overlapped peaks in powder diffraction patterns \cite{billinge2008powder_diffraction} and this problem is greatly amplified in  nano-materials, defined as crystallites with dimensions below 1000~\AA\  \cite{nano_billinge2007problem}, because the Bragg peaks become significantly broadened due to finite size effects.
Our goal is to see whether the use of prior knowledge in the form of previously solved structures, used to train a generative AI model, can overcome this challenge of structure solution from information-degraded diffraction patterns.

% What people have tried to do: traditional
%In general, no closed-form solution exists and this is an iteratively solved inverse problem.
%Depending on the complexity of the structure and the (degraded) information content of the signal, this inverse problem may or may not be poorly-constrained. 
In practice, for sufficiently simple structures it is possible to solve a structure from powder diffraction data alone \cite{billinge2008powder_diffraction, egami2012underneath}. Unsurprisingly, in the past year, a few concurrent works have tackled this setting of the problem for \textit{non}-nanoscale materials \cite{riesel2024crystal, li2024powder, lai2024end}.
%When the crystallite size becomes nano, there is further information loss in the powder diffraction signal due to Bragg peak broadening and overlap.
Yet, in the case of nano-materials it is rarely possible to solve the structure \textit{ab initio} -- we tackle this previously \textit{unsolved} problem \cite{juhas2006ab, juhas2010crystal}.

We use as input a powder x-ray diffraction (PXRD) pattern, measured over a limited range of momentum transfer, $Q$, with peak broadening due to crystallite nano-sizing. 
We note that,  
once demonstrated, the approach can also be applied to powder patterns that have no finite-size broadening, resulting in a broadly applicable method for structure solution from powder and nano-powder diffraction data.
The method shows great promise despite the highly degraded information content in the input patterns.

% What people have tried to do: modern auxiliary solutions
Although there are a number of works applying ML to problems in structure solution, we are not aware of any on par with our approach in breadth of applicability \cite{survey_billinge2024machine, survey_choudhary2022recent}.
Much of the early work in this area used machine learning to predict isolated properties such as unit cell parameters, magnetism, phase composition and symmetry group of structures given PXRD patterns \cite{xrd_class_oviedo2019fast, xrd_class_ml_suzuki2020symmetry, xrd_park2017classification, xrd_lee2020deep, xrd_aguiar2019decoding, class_ziletti2018insightful, xrd_tiong2020identification, garcia2019learning, merker2022machine, maffettone2021crystallography, szymanski2021probabilistic, uryu2024deep, szymanski2023autonomous, science_larsen2024phai}. 
We also note that our problem is distinct from that of structure prediction, where generative artificial intelligence (AI) techniques have been used successfully to generate novel (never seen before) material structures that are presumably energetically stable \cite{jiao2024crystal, pakornchote2024diffusion, antunes2023crystal, zhao2023physics}, \textit{e.g.}, GNoME \cite{gnome_merchant2023scaling}, UniMat \cite{unimat_yang2023scalable}, CDVAE \cite{cdvae_xie2021crystal}, MatterGen \cite{zeni2023mattergen}. 
% What people have tried to do: direct solutions and why they don't work
ML has also been used to help with sub-tasks in the overall structure solution problem, such as structure solution for a limited number of of crystal systems (\textit{CrystalNet} \cite{crystalnet_guo2023towards}) and close-packed structures (\textit{DeepStruc} \cite{kjaer2023deepstruc}), AI accelerated mining of existing structural databases (\textit{MLstructureMining} \cite{kjaer2024mlstructuremining}), single-crystal structural solutions (\textit{Crysformer} \cite{dun2023crysformer, pan2023deep}), or structure solution with single-element materials \cite{kwon2023spectroscopy}. 

Furthermore, in the past year, a few concurrent works employ generative AI approaches to solve structures end-to-end from PXRD \cite{li2024powder, lai2024end, riesel2024crystal}. However, the crucial limitation of all these prior works is that they only solve PXRD patterns \textit{without} finite-size effects. As previously noted, structure determination in this non-nanoscale setting is considered by materials scientists to be more or less a solved problem \cite{billinge2008powder_diffraction, egami2012underneath}. On the other hand, structure determination from nanoscale materials is significantly harder from a physical perspective, due to peak broadening caused by the finite crystal size \cite{juhas2008liga, juhas2010crystal, juhas2006ab}. Another major shortcoming of these works is that their code is not (yet) publicly available, thereby limiting reproducibility and/or comparisons to their work.

% What we hope to improve on
We find that our use of diffusion models for a fully end-to-end structure solution tool is highly promising. It was able to find plausible structures from information degraded PXRD patterns across broad classes of materials from all seven crystal systems. It was designed to output the lattice parameters and fractional atomic coordinates of atoms in a unit cell of the structure, in a crystallographic setting that can be ingested by other programs that take crystal structures as inputs.

% Work that inspires us to a solution

% What we do
%We propose \nanogen, an end-to-end graph diffusion model that takes as input the unit cell chemical formula and PXRD pattern (in principle, the diffraction pattern could be from any source), and outputs structure solutions with lattice parameters and atomic coordinates. 
Owing to the inherent uncertainty when presented with information-scarce diffraction patterns, \nanogen, via its Langevin dynamics-inspired conditional generative process \cite{song2019generative}, generates multiple candidate structures that adhere to the input information. 
These candidates can then be refined using local search methods, finding the global minimum structure solution if the model places the candidate anywhere within its basin of attraction.
It successfully found structure solutions for simulated nanocrystals as small as 10~\AA, is successful across all seven crystal systems, and was shown to work on real-world experimentally gathered PXRD patterns.

To further community research efforts in structure solution, we also propose \mppxrd, a benchmark task of inorganic materials and their corresponding simulated PXRD patterns. The data itself is curated from the Materials Project \cite{cdvae_xie2021crystal, materials_project_jain2013commentary}, and our hope is that it provides a uniform standard of comparison for future computational materials scientists working on structure solutions from PXRD. Its advantages (aside from the PXRD information) are that the materials in the dataset are generally stable, experimentally synthesizable, and diverse, while containing at most 20 atoms in the unit cell \cite{cdvae_xie2021crystal}. It is freely downloadable at this link: \href{https://github.com/gabeguo/cdvae_xrd/tree/main/data/mp_20}{https://github.com/gabeguo/cdvae\_xrd/tree/main/data/mp\_20}.

% % Summary
% In summary, our contributions are:
% \begin{enumerate}
%     \item We introduce \nanogen, a conditional graph diffusion model that is, to our knowledge, the first in the world to automatically generate plausible end-to-end structure solutions for nanomaterials, given standard PXRD patterns and chemical formulas.
%     \item We find that \nanogen can solve nanostructures of width as small as 10~\AA, with up to $50.0\%$ of its predictions matching the ground truth, and the average prediction (including non-matches) having $79.8\%$ overall structural similarity with the ground truth, based on our simulated nanoscale shrinkage experiments. 
%     \item We validate that \nanogen is successful on real-world PXRD patterns, making correct structural predictions $40.0\%$ of the time on our real-world dataset, with average $81.5\%$ overall structural similarity with the true structure.
%     \item We demonstrate that \nanogen's structure solution ability is consistent across all seven crystal systems; the best result is that it found the correct structure for cubic crystals $80.8\%$ of the time.
%     \item We introduce \textit{MP-20-PXRD}, a benchmark dataset of inorganic materials and simulated PXRD patterns, for the community to evaluate the performance of methods that generate structure solutions from PXRD conditioning.
% \end{enumerate}

\begin{table}[t]
    \centering
    \begin{tabular}{l | l l | r}
        \rowcolor[gray]{0.85} \textbf{Data Source} & \textbf{\nanogen} & \textbf{CDVAE-Search} & \textbf{Number} \\
        \rowcolor[gray]{0.85} & \textbf{(Ours)} & \textbf{(Baseline)} & \textbf{Materials} \\
        \hline
        MP-20 \text{ }\text{ }\text{ } (10~\AA) & $\mathbf{0.325} \pm 0.283$ & $0.647 \pm 0.301$ & $200$ \\
        MP-20 \text{ } (100~\AA) & $\mathbf{0.324} \pm 0.261$ & $0.591 \pm 0.273$ & $200$ \\
        Experimental & $\mathbf{0.383} \pm 0.330$ & $0.634 \pm 0.311$ & $15$ \\   
    \end{tabular}
    \caption{$\mathbf{R_{wp}^{2}}$ \textbf{(Material Reconstruction Success):} $\mathbf{R_{wp}^{2}}$ (otherwise known as the \textit{R-Factor} or \textit{Residuals Function}) refers to the (pre-Rietveld refinement) relative error between the PXRD pattern of the ground truth material and the PXRD pattern of the predicted material, where lower values are better \cite{egami2012underneath}; we calculate the mean and standard deviation over the testing set. \textit{Data Source} refers to the source of the data, where "MP-20 ($\{10, 100\}$~\AA)" 
    refers to PXRD patterns of simulated $\{10, 100\}$~\AA-scale nanomaterials from the MP-20-PXRD dataset \cite{materials_project_jain2013commentary, cdvae_xie2021crystal}, and "Experimental" refers to PXRD patterns from real-world experimental data. 
    ``CDVAE-Search" refers to results from latent space exploration of the unconditional CDVAE diffusion model.
    } \label{tab:results}
\end{table}

\begin{figure}
    \centering
    \includegraphics[width=\textwidth]{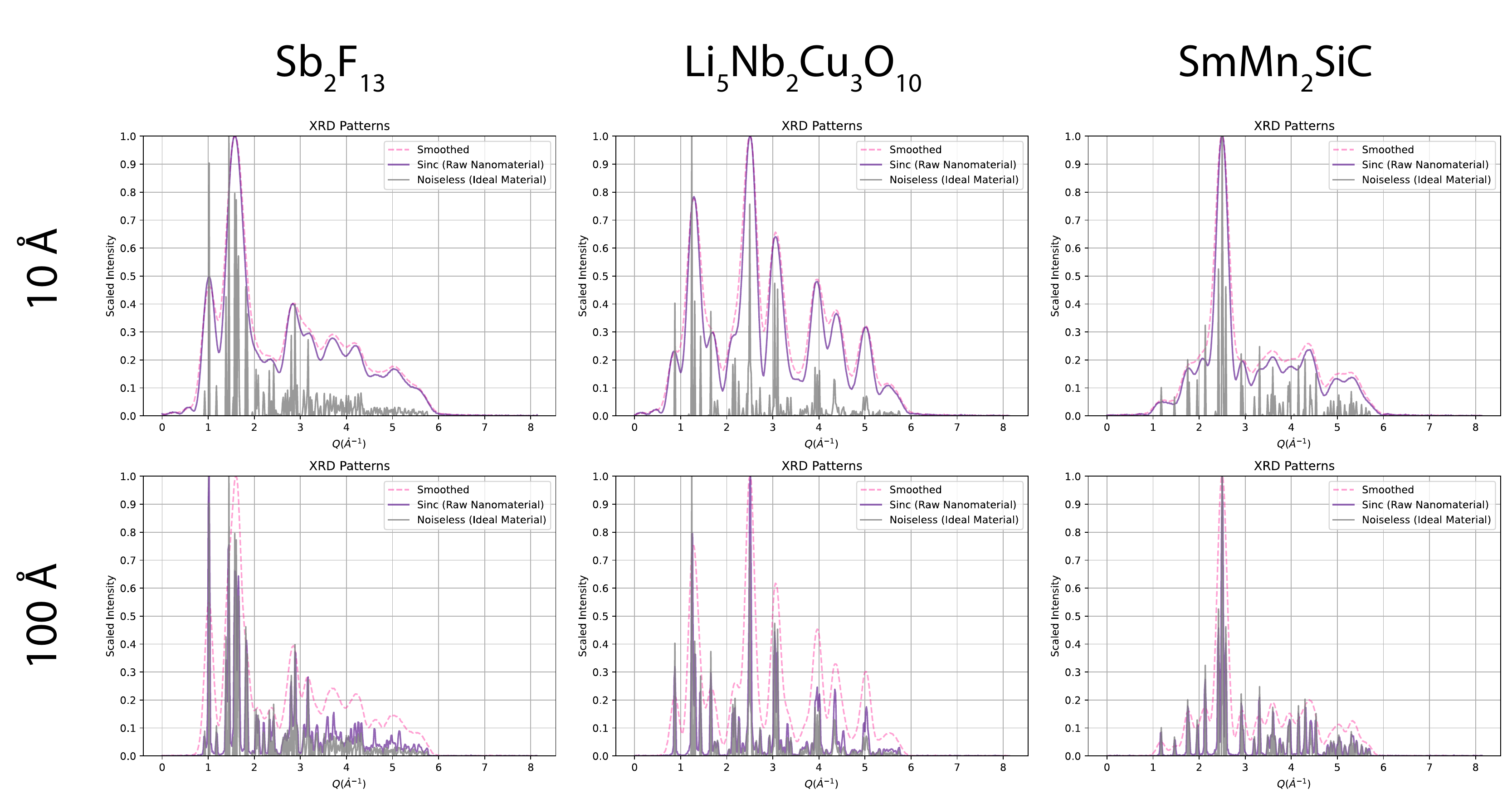}
    \caption{\textbf{{Nanomaterial PXRD Patterns}:} We simulate nanoscale shrinkage via the sinc$^2$ filter, thereby broadening the PXRD peaks (purple lines) from the ideal pattern (gray lines). To improve model performance, we create a smoother target PXRD pattern (dotted pink lines) that removes the sharp ripples at the expense of further broadening the diffraction pattern.  This is done via an additional Gaussian filter after the sinc filter. The horizontal axis is $Q$ (\AA$^{-1}$), and the vertical axis is scaled intensity (where $1$ is maximal).}
    \label{fig:pxrd_patterns}
\end{figure}

\begin{figure}
    \centering
    \begin{subfigure}[h]{0.48\textwidth}
        \includegraphics[width=\textwidth]{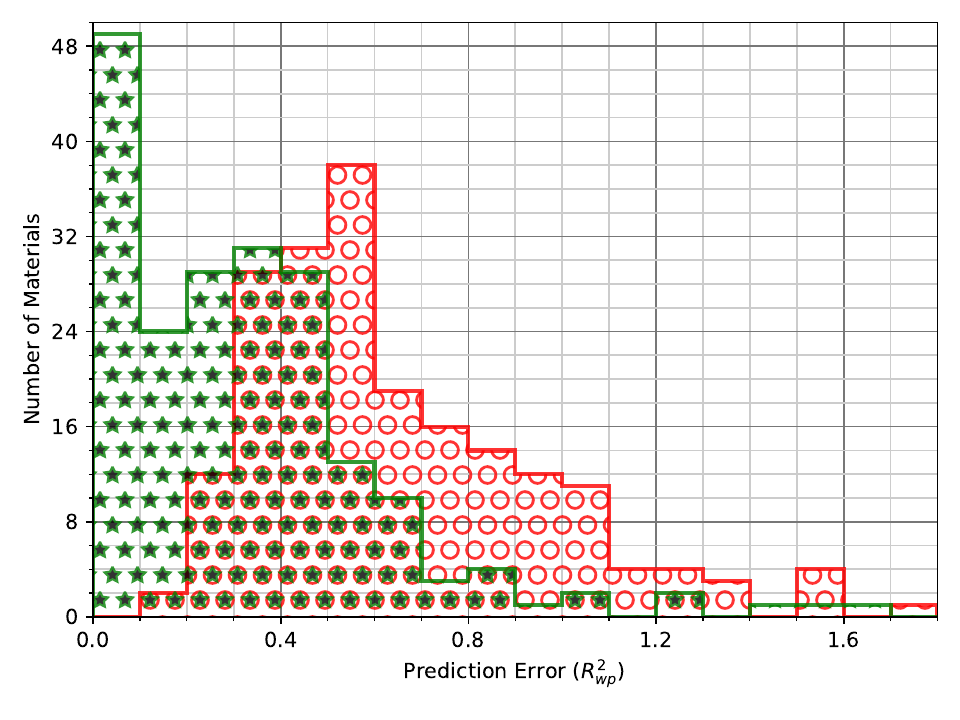}
        \caption{10~\AA}
    \end{subfigure}
    \begin{subfigure}[h]{0.48\textwidth}
        \includegraphics[width=\textwidth]{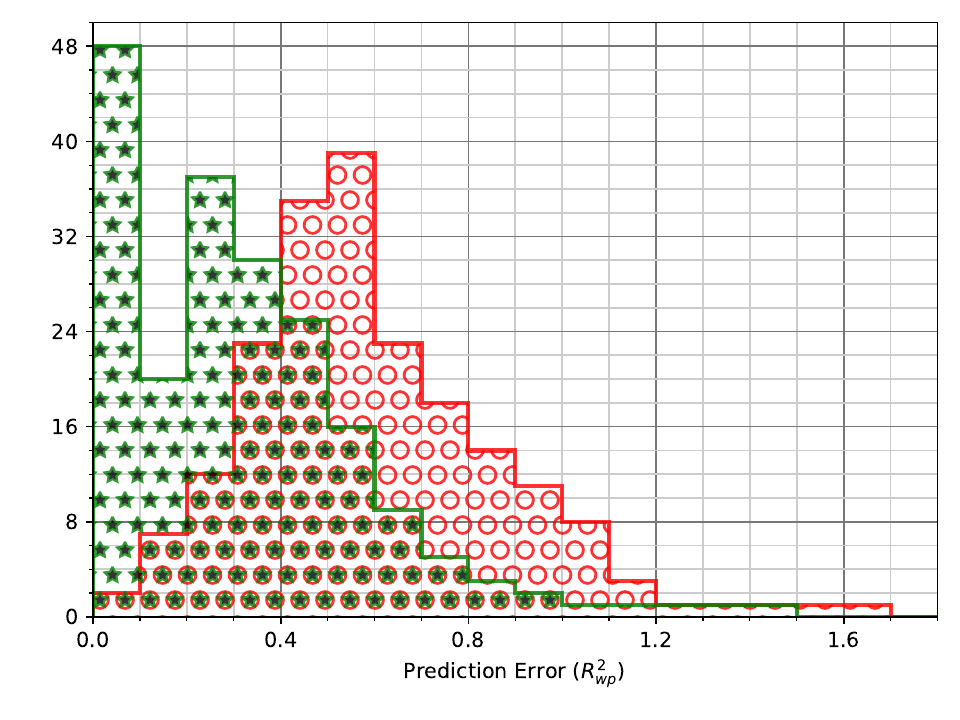}
        \caption{100~\AA}
    \end{subfigure}
    \caption{\textbf{Histogram (Distribution) of $\mathbf{R_{wp}^{2}}$:} Corresponds to \rw results shown in Table~\ref{tab:results}. The horizontal axis is \rw (the R-factor), \textit{i.e.}, the relative error between the predicted material's PXRD pattern and the ground truth material's PXRD pattern. The vertical axis is the number of materials from the testing set that had \rw at a certain level. The green line with stars represents \nanogen candidates, and the red line with dots represents CDVAE-Search candidates. Panel A is from PXRD patterns from 10~\AA\ nanomaterials; Panel B is from PXRD patterns from 100~\AA\ nanomaterials.}\label{fig:r_histogram}
\end{figure}

% \begin{figure}
%     \centering
%     \begin{subfigure}[h]{0.48\textwidth}
%         \includegraphics[width=\textwidth]{r_value_plots/sinc10_r_value_cdf.pdf}
%         \caption{10~\AA}
%     \end{subfigure}
%     \begin{subfigure}[h]{0.48\textwidth}
%         \includegraphics[width=\textwidth]{r_value_plots/sinc100_r_value_cdf.pdf}
%         \caption{100~\AA}
%     \end{subfigure}
%     \caption{\textbf{Cumulative Distribution Function of R-factors}: Horizontal axis is the R-factor, \textit{i.e.}, the relative error between the \nanogen-predicted material's PXRD pattern and the ground truth material's PXRD pattern. Vertical axis is the percentage of materials in the testing set that have R-factor at or below a certain level. Lower R-factors are better. Panel A is from PXRD patterns from 10~\AA\ nanomaterials; Panel B is from PXRD patterns from 100~\AA\ nanomaterials.}\label{fig:r_cdf}
% \end{figure}

\section*{Results}

%\subsection*{\nanogen Succeeds under Simulated Nanoscale Shrinkage}

\subsection*{Nanostructure Evaluation Setup}

Typically, a nanostructure is defined as that with crystalline size below 1000~\AA\ \cite{nanomaterial_definition_kreyling2010complementary}. To test the efficacy of our approach, we go two orders of magnitude below that and simulate PXRD patterns from crystalline sizes 10~\AA\ and 100~\AA\ with a mathematically principled filtering approach based on Fourier analysis \cite{lathi2005linear, fourier_pxrd_broad_sivia2011elementary} (described in Methods). 
As expected, the 10~\AA\  case has more peak broadening (and therefore, information degradation) \cite{nano_billinge2007problem, billinge2008powder_diffraction} than the 100~\AA\ case. 
%While the 10~\AA\ case has more peak broadening than the 100~\AA\ case, we notice that the $1000 \text{~\AA}$ case actually exhibits the most peak broadening. This is due to undersampling, as defined by the Nyquist sampling theorem \cite{lathi2005linear}. We deliberately pick this $1000 \text{~\AA}$ example to show the robustness of our method, even under the limits of digital signal processing. More details are available in Methods.
Visualizations of representative PXRD patterns are shown in Figure~\ref{fig:pxrd_patterns}.

To our model, we input this nanoscale PXRD pattern and the chemical formula of materials from MP-20 (a subset of the Materials Project database) \cite{cdvae_xie2021crystal, materials_project_jain2013commentary}. The model outputs a set of material structure predictions in the form of lattice parameters and fractional coordinates of atoms in the unit cell. We have trained separate models for each nanocrystalline size, for a total of two models. 
We also have CDVAE-Search \cite{cdvae_xie2021crystal} as our baseline model (which is, in some sense, an ablation on \nanogen): this searches and decodes a trained diffusion model's latent space for materials that adhere to the PXRD pattern, \textit{without} using any PXRD conditioning to generate the initial latent codes. We emphasize that at the moment, competing generative AI works for structure solution do \textit{not} make their source code public, nor do they handle finite-size effects, thereby making CDVAE-Search the closest comparison we have.
See Methods for more details.

See Figure~\ref{fig:visualizations} for representative structure predictions, Figure~\ref{fig:pxrd_comparison} for the corresponding PXRD patterns, and Table~\ref{tab:results} for success metrics.
In reviewing our results, we emphasize that the nanostructure problem currently has no end-to-end solution in the literature.

\subsection*{Qualitative Analysis}

\begin{figure}
    \centering
    \includegraphics[width=\textwidth]{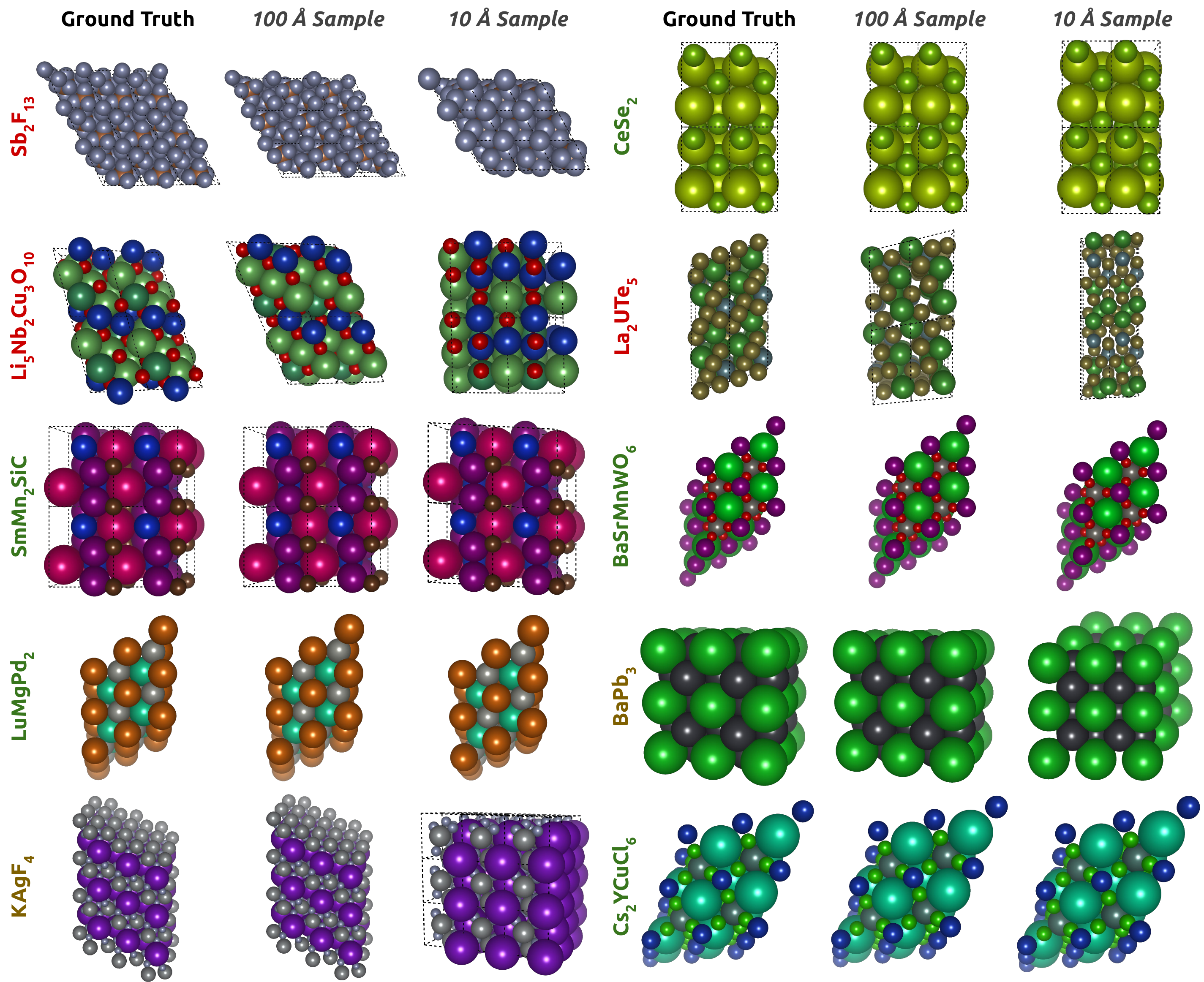}
    \caption{\textbf{\nanogen Structure Predictions}: The leftmost column is the ground truth crystal structure. The other columns show \nanogen's reconstructed crystal structures (after Rietveld refinement) from simulated nanocrystal PXRD patterns of diameter 10, 100~\AA. For convenient visualization of some examples, we have tessellated their unit cells into super cells and translated the origin of the \texttt{cif} files. These structures have been \textit{uniformly selected} from the testing dataset. Thus, the spread of success and failure across these figures is roughly reflective of success and failure across the whole testing dataset.
    } \label{fig:visualizations}
\end{figure}

Figure~\ref{fig:visualizations} shows a representative sampling of post-Rietveld refinement \cite{billinge2008powder_diffraction} \nanogen outputs. We display the post-Rietveld versions because it is standard procedure in crystallography to do a local minimization refinement step, such as Rietveld refinement in the case of powder data, on structure candidates \cite{billinge2008powder_diffraction, david2006structure}.
We selected the solutions to put into Figures~\ref{fig:visualizations} by uniformly sampling materials from \nanogen's testing set. Thus, the spread of success and failure across these figures is roughly reflective of success and failure across the whole testing dataset. Among the refined candidates for each material, we display the one with the lowest post-refinement R-factor (in a real-world experimental setting, this is knowable, since we have the ground truth PXRD pattern) \cite{billinge2008powder_diffraction}.

We see \nanogen can succeed across materials with a wide variety of inorganic chemical compositions, ranging from \ce{BaSrMnWO6} to \ce{LuMgPd2}. 
We also see that performance is slightly better for 100~\AA\ simulated crystalline size, although 10~\AA\ is up-to-par in most cases. 
In particular, \nanogen is generally successful at capturing diverse lattice shapes, such as those of \ce{Cs2YCuCl6} and \ce{SmMn2SiC}. 
Furthermore, we see that \nanogen can capture symmetries in cases like \ce{Cs2YCuCl6} and \ce{BaSrMnWO6}. 
Even in failure cases like \ce{Li5Nb2Cu3O10} or \ce{Sb2F13}, \nanogen is still able to lead us towards reasonable atomic packing and lattice shapes.
% We note that our model is not learning symmetries present in the structure, but rather relative atomic positions in a unit cell, and so the concept of special positions is lost and the model can place the origin of the structure anywhere in the unit cell resulting in an equivalent structure.

\subsection*{Quantitative Analysis}

There is not a universally agreed upon definition of success for a structure solution. 
This fact has raised challenges in the field of structure prediction, where there is discussion about how many of millions of predicted ``new" structures \cite{gnome_merchant2023scaling} are actually novel \cite{cheetham2024artificial}, and likewise \cite{leeman2024challenges} for the results of robotic high throughput syntheses \cite{szymanski2023autonomous, szymanski2021toward}.  
This also presents a challenge for us to determine quantitatively the structures that are correct, mostly correct, or incorrect. 
This is made all the more difficult because structures are solved just up to isometry from powder diffraction data, and may be rotated, displaced or inverted compared to ground-truth but nonetheless correct. 

That being said, the crystallographic community generally uses \rw, the profile-weighted R-factor, as the success metric. This is the intensity-scaled sum of squares of the deviation of the predicted and ground truth powder patterns (see "Methods" for details). It is commonly used as a target function in Rietveld structure refinements. Lower values are better, with the minimum value being $0$, and \rw below $10\%$ generally considered successful \cite{billinge2008powder_diffraction, egami2012underneath}. For each material, we report the single best \rw out of \nanogen's top~$10$ returned candidates; this is reasonable, as in real-world scenarios, the ground-truth PXRD is measurable.

% In the field of structure solution from powder diffraction, a final structure refinement step is applied where a local search is carried out such that structural parameters (lattice parameters and fractional coordinates) are modified using non-linear regression methods to obtain the best fit to the measured powder diffraction pattern \cite{david2006structure}.
% Finally, the chemical plausibility (\textit{e.g.,} interatomic distances, bond angles, missing or extra sites in organic compounds) of the resulting structure is tested using CIF-checkers, most notably IUCr's CheckCIF \cite{check_cif}.
% These steps are not easily automated and are not high throughput, and so not appropriate for our purposes.  They are also problematic for us because our \textit{ab initio} approach does not first find the unit-cell with high accuracy as is done in the indexing process with high resolution powder diffraction methods.  
% We tried the local refinement and found that the local search is very sensitive to having correct unit cell parameters, even when these are less relevant parameters for small nanoparticles that are our target here.
% We believe that these issues will be resolved in the future, but rather than address them here we turn to other measures of success that are high throughput.
% We note that these may not accurately reflect the actual degree of success of our method.
% They are more likely to result in false negatives than false positives and likely under-count the success rate, though this needs to be established.
% The metrics are described in the "Methods" section.

The quantitative results of Table~\ref{tab:results} support the qualitative observations and show that \nanogen is  successful at reconstructing materials in MP-20. 
We see that \nanogen's predictions are considerably better than the CDVAE-Search baseline (more details in ``Methods"). 
The two \nanogen models perform at essentially the same level pre-Rietveld refinement, and they both considerably outperform the CDVAE-Search baseline. 
We note that before model refinement, even from conventional structure solution methods from well crystallized samples, the \rw measure can be considerably higher than 10\%, even for the correct structure.
% Furthermore, we observe that if we go one standard deviation below the mean, for the 100~\AA\ case, we barely hit the success threshold; for the 10~\AA\ case, we barely miss the success threshold. 
This shows that our method, while far from perfect, can give decent structural candidates for a considerable number of materials.

Figure~\ref{fig:r_histogram} shows the distribution of \rw for \nanogen (green), the latent search baseline (blue), and the random search baseline (red). We see that for both \nanogen models, the peaks of the distributions are concentrated around relatively low \rw, many below 10\%, with rightward tails representing higher \rw. In contrast, the CDVAE-Search baselines peak at higher \rw, with very few of their \rw crossing the 10\% threshold. Similarly to the \nanogen distribution, the latent search distribution is also somewhat right-skewed. This highlights the advantages of \nanogen specifically in generating high-quality structural candidates.

% Figure \ref{fig:r_cdf} shows us the cumulative distribution function of R-factors. We see that in both the 10~\AA\ and 100~\AA\ cases, almost $20\%$ of the predictions hit the success threshold. 

%Furthermore, in the future, we can use techniques like Rietveld refinement \cite{billinge2008powder_diffraction} to start from our already-plausible structure guesses and refine them (based on physical chemistry principles and the PXRD patterns) to near-exact structure solutions.

\begin{figure}
    \centering
    \includegraphics[width=\textwidth]{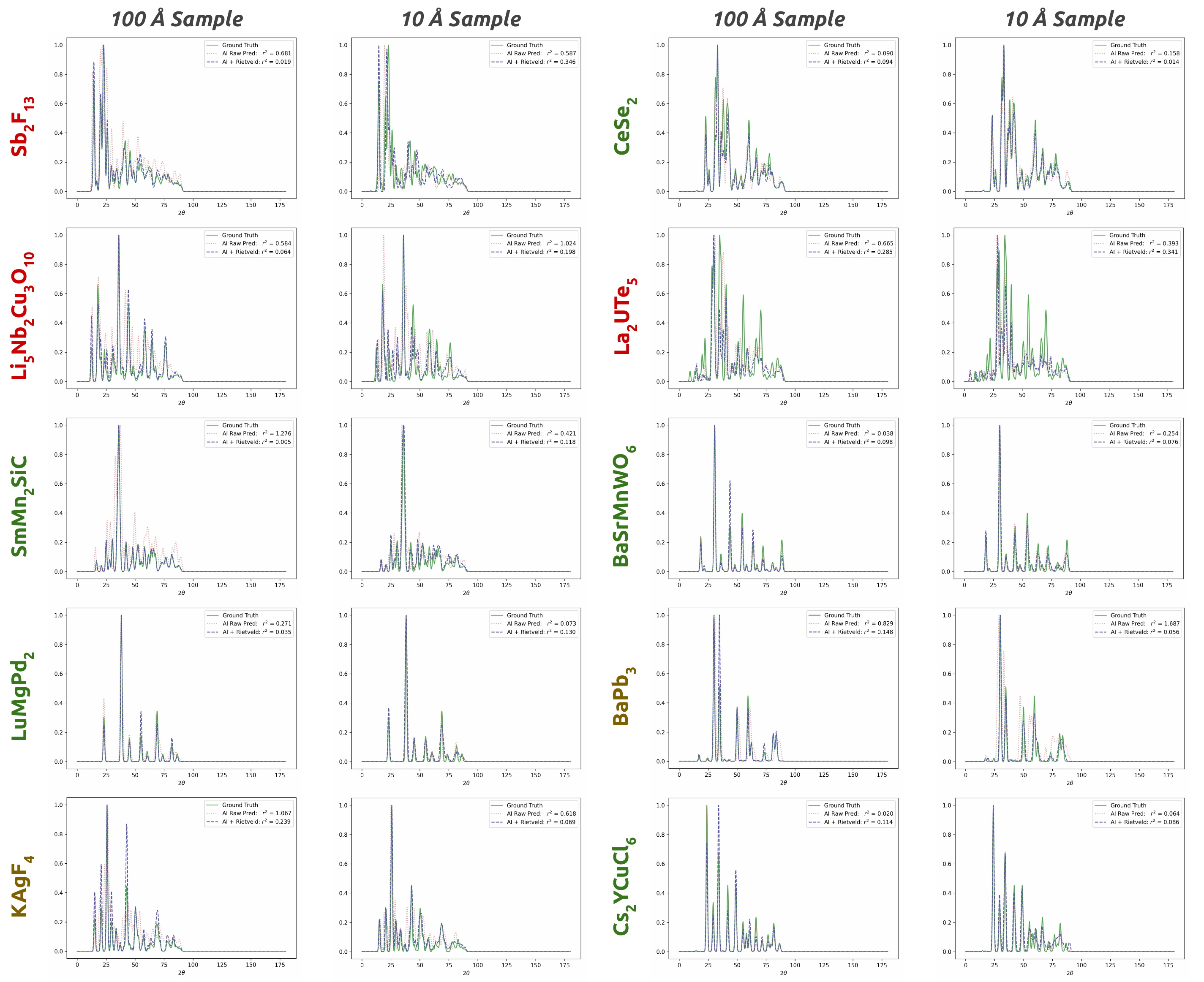}
    \caption{\textbf{PXRD Comparisons}: Comparison (without finite size effects) of the ground truth PXRD pattern, the raw \nanogen prediction's PXRD pattern, and the Rietveld-refined PXRD pattern. Corresponds to results in Figure \ref{fig:visualizations}.} \label{fig:pxrd_comparison}
\end{figure}

\begin{table}
    \centering
    \begin{tabular}{r | c c c c}
        \rowcolor[gray]{0.85} \textbf{Crystal System (\#)} & \textbf{\nanogen \text{ } (10~\AA)} & \textbf{\nanogen (100~\AA)} & \textbf{Latent Search (10~\AA)} & \textbf{Latent Search (100~\AA)}\\
        \hline
        \rowcolor[gray]{0.95} \multicolumn{5}{l}{$\mathbf{R_{wp}^{2}}$ \textbf{ ($\downarrow$)}}  \\
        Cubic ($52$) & $0.185 \pm 0.281$ & $0.224 \pm 0.311$ & $0.671 \pm 0.339$ & $0.650 \pm 0.344$ \\
        Tetragonal ($31$) & $0.443 \pm 0.315$ & $0.378 \pm 0.218$ & $0.685 \pm 0.268$ & $0.630 \pm 0.243$ \\
        Orthorhombic ($35$) & $0.301 \pm 0.208$ & $0.380 \pm 0.263$ & $0.636 \pm 0.296$ & $0.500 \pm 0.211$ \\
        Trigonal ($23$) & $0.388 \pm 0.228$ & $0.344 \pm 0.256$ & $0.678 \pm 0.220$ & $0.615 \pm 0.218$ \\
        Hexagonal ($12$) & $0.416 \pm 0.155$ & $0.366 \pm 0.195$ & $0.510 \pm 0.192$ & $0.542 \pm 0.152$ \\
        Monoclinic ($37$) & $0.318 \pm 0.160$ & $0.328 \pm 0.226$ & $0.607 \pm 0.307$ & $0.564 \pm 0.272$ \\
        Triclinic ($10$) & $0.547 \pm 0.521$ & $0.368 \pm 0.104$ & $0.679 \pm 0.383$ & $0.590 \pm 0.263$ \\
    \end{tabular}
    \caption{
    \textbf{Results by Crystal System:}
    \textit{Crystal System (\#)} refers to the crystal system name, and the number of materials we have in the dataset of that system. Interpretation of metrics and method names remains the same as Table \ref{tab:results}. All results shown are for MP-20.} \label{tab:results_by_crystal_system}
\end{table}

\subsection*{Comparison Across Crystal Systems}

Especially with information diminished inputs, more complex structures are harder to solve. This was seen in the recent study where  \textit{CrystalNet}, a generative AI encoder-decoder, was considerably more successful on simpler materials from cubic crystal systems \cite{crystalnet_guo2023towards}. To investigate this in the current case, the performace of \nanogen is broken out by  crystal system in Table~\ref{tab:results_by_crystal_system}, where the results are from the same experimental run as Table~\ref{tab:results}.

Unsurprisingly, and in line with previous work \cite{crystalnet_guo2023towards}, we find that the cubic system is the easiest to solve. 
This trend holds, regardless of nanomaterial crystal size. 
The success does decrease with decreasing symmetry of the structure, with the triclinic and tetragonal crystal systems performing poorly.

\subsection*{Test of structure refinement step}

\begin{figure}[t]
    \centering
    \begin{subfigure}[h]{0.48\textwidth}
        \includegraphics[width=\textwidth]{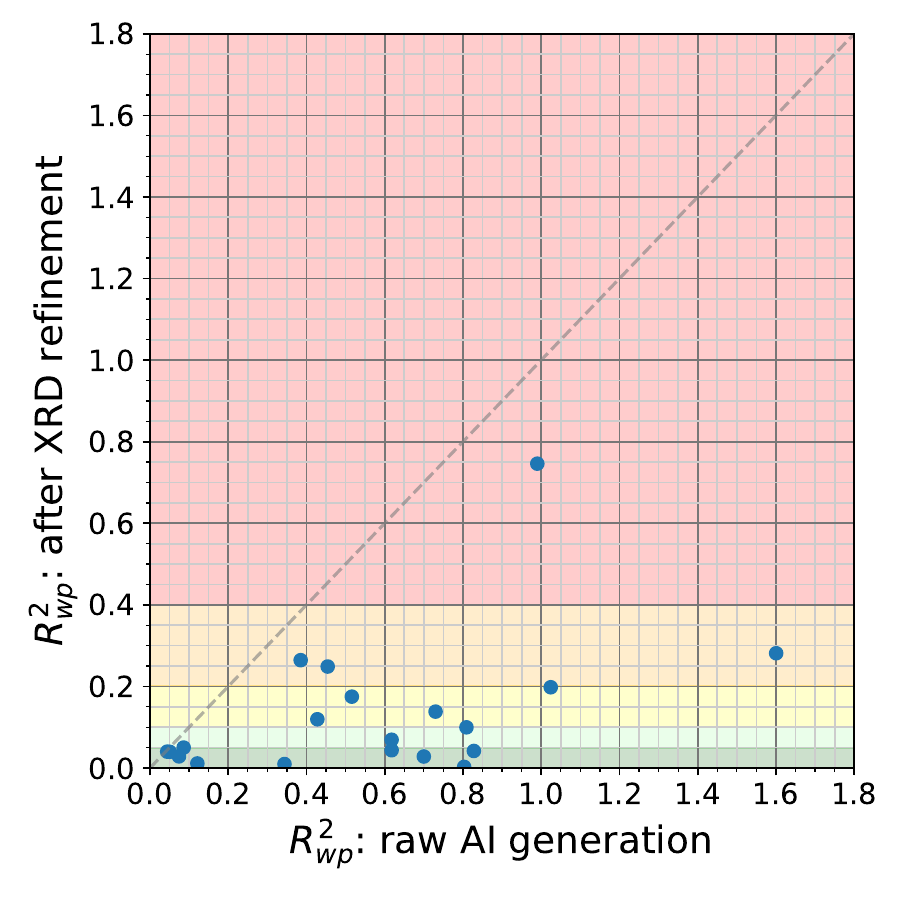}
        \caption{10~\AA.}
    \end{subfigure}
    \begin{subfigure}[h]{0.48\textwidth}
        \includegraphics[width=\textwidth]{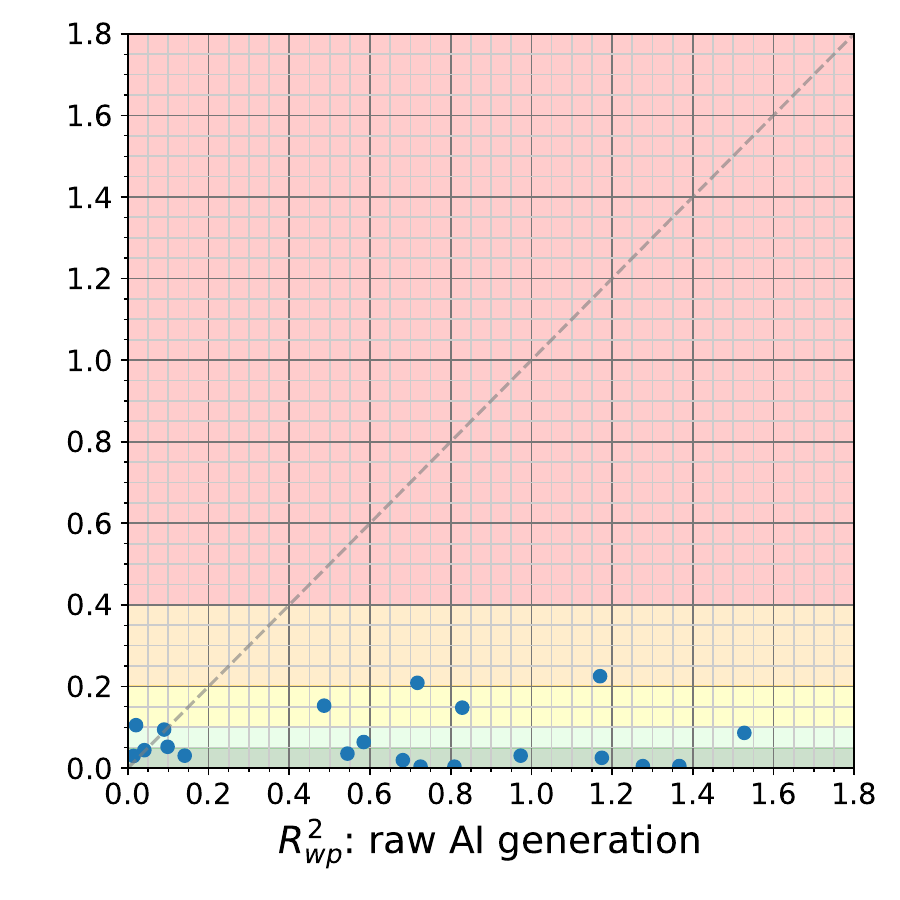}     
        \caption{100~\AA.}
    \end{subfigure}
    \caption{
    \textbf{Rietveld Refinement Results:} We conducted Rietveld refinement on ten promising candidates each for $20$ uniformly selected materials, making for a total of $200$ structures refined at each nanocrystal size. 
    Panel A shows the results for 10~\AA\ nanocrystal size; Panel B shows the results for 100~\AA\ nanocrystal size. We display the results corresponding to the best post-refinement \rw (which can be calculated without knowledge of the ground truth structure). The horizontal axis is the calculated \rw before refinement; the vertical axis is the calculated \rw after refinement. For 10~\AA\ nanocrystal size, the pre-refinement \rw has $\mu \pm \sigma = 0.561 \pm 0.388$, and the post-refinement \rw has $\mu \pm \sigma = 0.132 \pm 0.166$. For 100~\AA\ nanocrystal size, the pre-refinement \rw has $\mu \pm \sigma = 0.663 \pm 0.473$, and post-refinement \rw has $\mu \pm \sigma = 0.068 \pm 0.066$.}\label{fig:Rietveld}
\end{figure}

To further improve the results, we attempted a standard Rietveld refinement on a uniformly selected subset of 20 structures solved by \nanogen (including those displayed in Figure~\ref{fig:visualizations}). For each structure, we fed \nanogen's ten highest-ranked candidates. The results displayed are for the best candidate per material post-Rietveld refinement (again, this is knowable, since the ground truth PXRD pattern is observed).

Figure~\ref{fig:Rietveld} shows \rw values for these candidates before and after the Rietveld refinement step, and Figure~\ref{fig:pxrd_comparison} shows the PXRD patterns pre- and post-refinement. 
The Rietveld refinement works very well for the 100~\AA\ test cases, which have sharper Bragg peaks. 
In almost all of the cases (18 out of 20), the refinement brought \rw below 20\%. In a majority of the cases (15 out of 20), the refinement brought \rw below 10\%.
There may still be some minor issues with the structure in these cases, but \nanogen is clearly and consistently returning structures that are quite similar to the true one and likely close enough to yield a correct structure with some human input in each case.

It is also fascinating that the pre-refinement \rw of the candidates that led to the best post-refinement results were so high. We see some cases where a candidate with initial R-factor above $100\%$ got refined into near-perfect matches.

The Rietveld refinements were less successful for the 10~\AA\ nanoparticles, which have highly distorted diffraction patterns.  
While the refinement was usually helpful, it did not necessarily result in successful structures.

This aspect of the work can be developed in the future, with better understanding of how to refine structural models in a more automated way, either against the PDF \cite{nano_billinge2007problem, pdf_farrow2009relationship} or against powder data, to obtain a higher structure solution rate.

\subsection*{Performance with Experimental PXRD Patterns of Real-World Materials}

Finally, we test on a small number of experimentally determined PXRD patterns.
The patterns were obtained from IUCr's database \cite{iucr_database}.
We did artificially induce sinc and Gaussian filters (\textit{i.e.}, simulated nanoscale) on these patterns before feeding them into our model, to make them consistent with the training data that did undergo such filters. See "Methods" for more details.

\begin{figure}
    \centering
    \includegraphics[width=\textwidth]{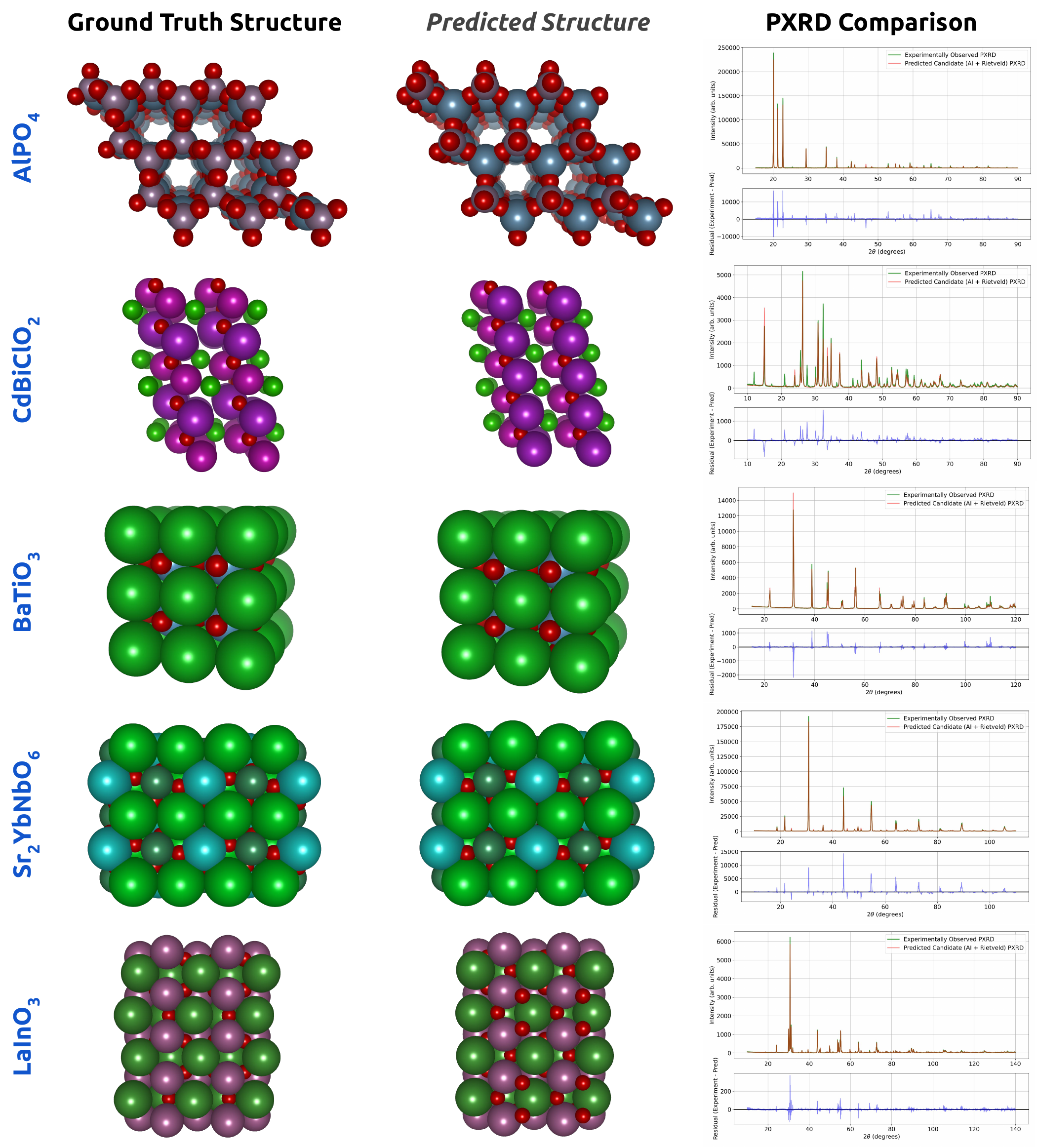}
    \caption{\textbf{Experimental Data}: Structure solutions from \textit{experimentally} observed PXRD patterns in the IUCr database \cite{iucr_database}. The leftmost column is the ground truth structure, the middle column is the predicted structure from \nanogen (lowest \rw candidate post-Rietveld refinement), and the rightmost column is the comparison of the simulated PXRD (red) from TOPAS v7 \cite{coelho2018topas} to the actual \textit{experimentally} observed PXRD (green).} \label{fig:experimental_success}
\end{figure}

Encouragingly, we find that \nanogen overcomes the simulation-to-real gap, in that its success, as measured by both visual analysis (Figure~\ref{fig:experimental_success}) and quantitative metrics (Table~\ref{tab:results}), are comparable to those obtained from the simulated data.

In particular, after Rietveld refinement in TOPAS v7 \cite{coelho2018topas}, we can achieve good adherence to the experimentally determined PXRD, and closely match the reported ground truth structure, as shown in Figure~\ref{fig:experimental_success}. There are some slight imperfections, however: in some cases, we see that the shape is correct, but the ``coloring" (\textit{i.e.}, species) of atoms is swapped, \textit{e.g.}, \ce{AlPO4}, \ce{CdBiClO2}. We also see cases where the structure is almost correct, but some atoms are placed slightly differently yet with some semblance of order -- for example, in both \ce{Sr2YbNbO6} and \ce{LaInO3}, the oxygen atoms are misplaced. 

We also note that the results displayed in Figure~\ref{fig:experimental_success} are from the best (lowest \rw) candidate post-Rietveld refinement, as measured by adherence to the experimentally observed PXRD -- other candidates may not be as successful. Nonetheless, this is a fair representation because in real scenarios, this PXRD is available to compare against.

\section*{Discussion}

The Diffusion model implemented in \nanogen provides an exciting step towards structure solution from information compromised input data such as would come from nanomaterials.  As an end-to-end method to solve nanostructures from information-compromised diffraction data, it is easy to use. 

As with any structure solution method, we do not expect 100\% success but seek a method that can provide structural candidates that may be further assessed for their validity. In this regard \nanogen has shown a high capability due to the inherent stochasticity of Langevin dynamics \cite{song2019generative},
which results in multiple structural candidates.

More work will be needed to understand exactly the ways that the structures fail, for example, in terms of incorrect geometry vs. chemical switches, and we believe that novel refinement methods can be developed that are better tuned to error correction in \nanogen generated structures than is Rietveld refinement.

We find it interesting that \nanogen actually works very well in cases where the information content was previously thought to be very low, \textit{e.g.}, 10~\AA\ nanocrystal size. This provides an optimistic note for future computational crystallographic and materials science research, as it shows that problems that are harder for humans are not necessarily harder for data-driven and machine learning methods.

Finally, we note that while we evaluated on PXRD data, in principle, models could be trained to work on diffraction data coming from other popular modalities like electron or neutron diffraction. 

\section*{Limitations and Future Work}

One limitation of our model is that it requires \textit{a priori} knowledge of the chemical formula. While this is a reasonable assumption in crystallography \cite{billinge2008powder_diffraction, egami2012underneath, crystalnet_guo2023towards}, it would be even more powerful if, given partial or even no information about the chemical composition, our model could still solve structures, and this will be explored. 
Another limitation of our current work is that materials from MP-20 have at most $20$ atoms in the unit cell \cite{cdvae_xie2021crystal}. This covers a wide range of useful materials, but many important materials have many more atoms in the unit cell. Scaling up our approach to materials with more atoms in the unit cell will also be the subject of future work.

On the other hand, the data quality that was used in this study was low.  
Powder patterns were used over a limited range to $Q<8.2$~\AA$^{-1}$ and broadened because of the nanoparticle sizing.  
In practice, it is often the case that real data available, for example, from x-ray synchrotron sources, has more information than the patterns that were used in this first test.
Training models that work on higher quality diffraction patterns can be expected to extend the quality of structure predictions from \nanogen.

From a machine learning perspective, we use an off-the-shelf CDVAE backbone \cite{cdvae_xie2021crystal} to demonstrate this proof of principle, sharing many design and architectural similarities with concurrent works \cite{li2024powder, lai2024end, riesel2024crystal} (although the others are not open-sourced). Thus, the novelty of our work is not in the machine learning advance, but rather in the fact that we chart a path to successfully solving the once-unsolved grand scientific challenge of structure determination of nanometer-sized materials.

The model is already out-performing in terms of structural complexity earlier attempts at nanoparticle structure solution~\cite{juhas2006ab, juhas2008liga, juhas2010crystal}, and more refinements to the approach are eminently possible.
One example of this is that our model is based on a CDVAE \cite{cdvae_xie2021crystal} backbone, but there have been recent developments in graph diffusion models \cite{jiao2024crystal, jiao2024space, linequivariant, luo2024towards} that could easily be swapped in for the CDVAE backbone.

Another future direction is to make the model robust to background signals (\textit{e.g.}, solvent, container). We fed our model experimental data from a colleague that had a significant signal from the sample container, which confused the model -- performance improved once we subtracted the container signal.

% Finally, in the future, we wish to conduct structural refinement as the final step of the evaluation pipeline. The ultimate goal of this line of research is to generate candidate material structures that can then be refined (via Ritveld refinement \cite{billinge2008powder_diffraction}, for instance) to the ground truth material that perfectly matches the PXRD pattern. It would be valuable to see the extent to which structural refinement improves the quality of our predictions, or even if it doesn't, perhaps due to some local minima.

\section*{Methods}

%\subsection*{Nanoscale Shrinkage} \label{sec: nanoscale shrinkage}
%Predicting nanomaterial structure from PXRD data is a fundamentally challenging problem. As the size of our crystal of interest falls, the uncertainty in our experimental observations of the crystal rises. This renders material structure discovery for nanomaterials very challenging, as experimental data only ever contains partial observations of the true material properties and structure. Interestingly, this corruption of information is an inherent physical phenomenon and isn't an artifact of the experimental process. 

%To understand the mechanism driving the information degradation, one must understand representations of molecule structure and their interplay. The pair distribution function (PDF) is a common representation of material structure, as it captures the distribution of relative distances between atoms in a crystal. In an experimental setting, however, materials are usually observed through and characterized by their PXRD pattern. Mathematically, these two representations contain the same information about the crystal. In fact, the ideal PXRD pattern is simply the Fourier Transform of the pairwise density function \cite{billinge2008powder_diffraction}.  

%The duality between the PDF and PXRD pattern of a material gives rise to inherent uncertainty in PXRD observations. Analyzing a small but representative portion of a crystal masks out portions of the PDF of the original crystal. 

\subsection*{Dataset}

\subsubsection*{Materials Project: MP-20-PXRD}

We used the MP-20 dataset of the Materials Project \cite{materials_project_jain2013commentary} which consists of materials sampled from the Materials Project database with at most 20 atoms in the unit cell \cite{cdvae_xie2021crystal}. Powder diffraction patterns for all structures in MP-20 were simulated using the PyMatGen package \cite{pymatgen_ong2013python}. Our simulation uses CuKa radiation with $Q$ values ranging from $0\le Q \le 8.1568$~\AA$^{-1}$.

The files are made available as a dataset we call MP-20-PXRD which we include with our open source code. 
We hope that this will encourage the computational material science community to extend our work and explore the nanostructure solution from PXRD problem even more.

We use MP-20-PXRD to train \nanogen end-to-end. 
MP-20-PXRD contains 45,229 materials, and we use a 90-7.5-2.5\% train-validate-test split. 

\subsubsection*{Experimental PXRDs}

We obtained $15$ experimental patterns from searching IUCr's database \cite{iucr_database}. 
The difficulty in procuring experimentally gathered PXRDs \textit{en masse} is that the reporting format for such patterns is not standardized. There is a powderCIF~\cite{pdcif_toby20064} format to standardize this but few authors deposit their powder data in this form, if at all.  Furthermore, even among the diffraction patterns available, many are from other scattering sources, like neutron diffraction. In future works, we plan to more systematically and extensively curate experimental PXRD patterns.
It would be very helpful if powder patterns were deposited in a standard format with published papers.

We also use the 100~\AA\ filter and Gaussian broadening on the experimental PXRD data, to make the patterns consistent with what was seen during train time by our model.

\subsection*{Finite size effects} \label{sec: nanoscale shrinkage}
% To model the nano-scale finite size of the particles we assumed the diffraction pattern to be that of a long-range ordered crystal, convoluted with the Fourier transform of a window function on the pair distribution function, 
% \begin{equation}\label{eqn:window}
% \Pi(r/\tau) = \begin{cases}
%       0 & |r| > \frac{\tau}{2} \\
%       \frac{1}{2} & |r| = \frac{\tau}{2} \\
%       1 & |r| < \frac{\tau}{2} \\
% \end{cases}
% \end{equation}
% where $r$ is the interatomic distance and $\tau$ is the nanomaterial width. We therefore convolute the diffraction pattern with its Fourier transform,
% \begin{equation}\label{eqn:sinc_filter}
% \mathcal{F}\left\{\Pi\left(\frac{r}{\tau}\right)\right\} = \tau\frac{\text{sin}(\frac{Q \tau}{2})}{\frac{Q \tau}{2}}
% \end{equation}
% where $Q$ is the magnitude of the scattering vector \cite{egami2012underneath}. 

To model the nano-scale finite size of the particles, we follow the approximation of a finite size particle as a periodic crystal windowed over a certain domain in real space. 
This does not capture non-space filling clusters, faceted morphologies, surface relaxations and so on, which we defer to future studies.
We then make use of Fourier analysis \cite{pdf_farrow2009relationship} to derive the correct form for the diffraction pattern of a nano-scale crystal.

A physically principled and computationally feasible approximate model is to convolute the diffraction pattern with a sinc$^2$ function \cite{fourier_pxrd_broad_sivia2011elementary}: 

\begin{equation}\label{eqn:conv_sinc2}
    I_\text{nano} = I_\text{c} \circledast \text{sinc}^{2}\left(\frac{Q\tau}{2}\right),
\end{equation}
where $Q$ is the magnitude of the scattering vector, $\tau$ is the size of the nanoparticle, $I_c$ is the diffraction pattern of the infinitely repeating crystal, and $I_\text{nano}$ is the diffraction of the nano-scale crystal. Intuitively, the sinc$^2$ kernel has the desirable properties: (a) it is non-negative, thereby being physically plausible; (b) as $\tau$ decreases (corresponding to smaller crystalline size), the kernel gets wider, thereby broadening the peaks further. In particular, keeping in mind the Fourier relationship between the interatomic distance and the diffraction pattern, finite size effects can be seen as multiplying the distribution of interatomic distances with the window function
\begin{equation}\label{eqn:window}
\Pi(r/\tau) = \begin{cases}
      0 & |r| > \frac{\tau}{2} \\
      \frac{1}{2} & |r| = \frac{\tau}{2} \\
      1 & |r| < \frac{\tau}{2} \\
\end{cases}.
\end{equation}
Then, going from real space to Fourier (\textit{i.e.}, diffraction pattern) space,
\begin{equation}\label{eqn:fourier_window}
    \mathcal{F}\left[\Pi(r/\tau)\right] = \tau \cdot \text{sinc} \left(\frac{Q \tau}{2}\right).
\end{equation}
The leading $\tau$ factor can be ignored, since intensity is on an arbitrary scale. 
%The eventual squaring of the sinc comes from the fact that the diffraction intensity is proportional to the squared Fourier transform of the interatomic distance distribution. \sjb{removing this since, as stated, it is not true.}

Mathematically, this is only an approximation, but nonetheless a good one that appears in standard texts on the subject \cite{fourier_pxrd_broad_sivia2011elementary}.
The result of the convolution can be seen in Figure~\ref{fig:pxrd_patterns}.

\subsubsection*{Practical Considerations of Nanoscale Shrinkage}

Simulating nanoscale shrinkage effects for \textit{larger} nanomaterial widths requires sampling from increasingly high frequency sinc functions. As a result, discrete sampling from these signals requires a large sampling rate to avoid any aliasing effects. To that end, we simulate PXRD patterns as scalar-valued functions over 4096 evenly sampled points in the range 0 to 8.1568~\AA$^{-1}$, resulting in a sampling rate of 502.16~\AA. By the Nyquist Sampling Theorem \cite{1697831}, this sampling rate limits the bandwidth of any signal of interest (that is, the sinc pattern) to 251.08~\AA\ to avoid aliasing effects. All the sinc functions we use have maximum frequency below this sampling rate.

The convoluted diffraction pattern is then downsampled to the 512 input samples expected by \nanogen. 

After simulating nanoscale shrinkage, we also apply Gaussian smoothing with $\sigma$ ranging uniformly between 0.081568~\AA$^{-1}$ and 0.0897248~\AA$^{-1}$. 
Empirically, we found this improved \nanogen performance -- we believe that the smoothing produces more continuous regression targets for the PXRD-regressor.

We train and test a separate model for each crystal size (10~\AA, 100~\AA).

%Unsurprisingly, these aliasing effects are present in our experiments for simulated nanomaterial size $1000\text{~\AA}$, and can be seen in Figure \ref{fig:pxrd_patterns}. The aliasing results in more than expected information degradation, and in fact makes the problem more challenging. 

\subsection*{Base Model: Hybrid Diffusion-VAE}\label{sec:cdvae}

\begin{figure}
    \centering
    \includegraphics[width=\textwidth]{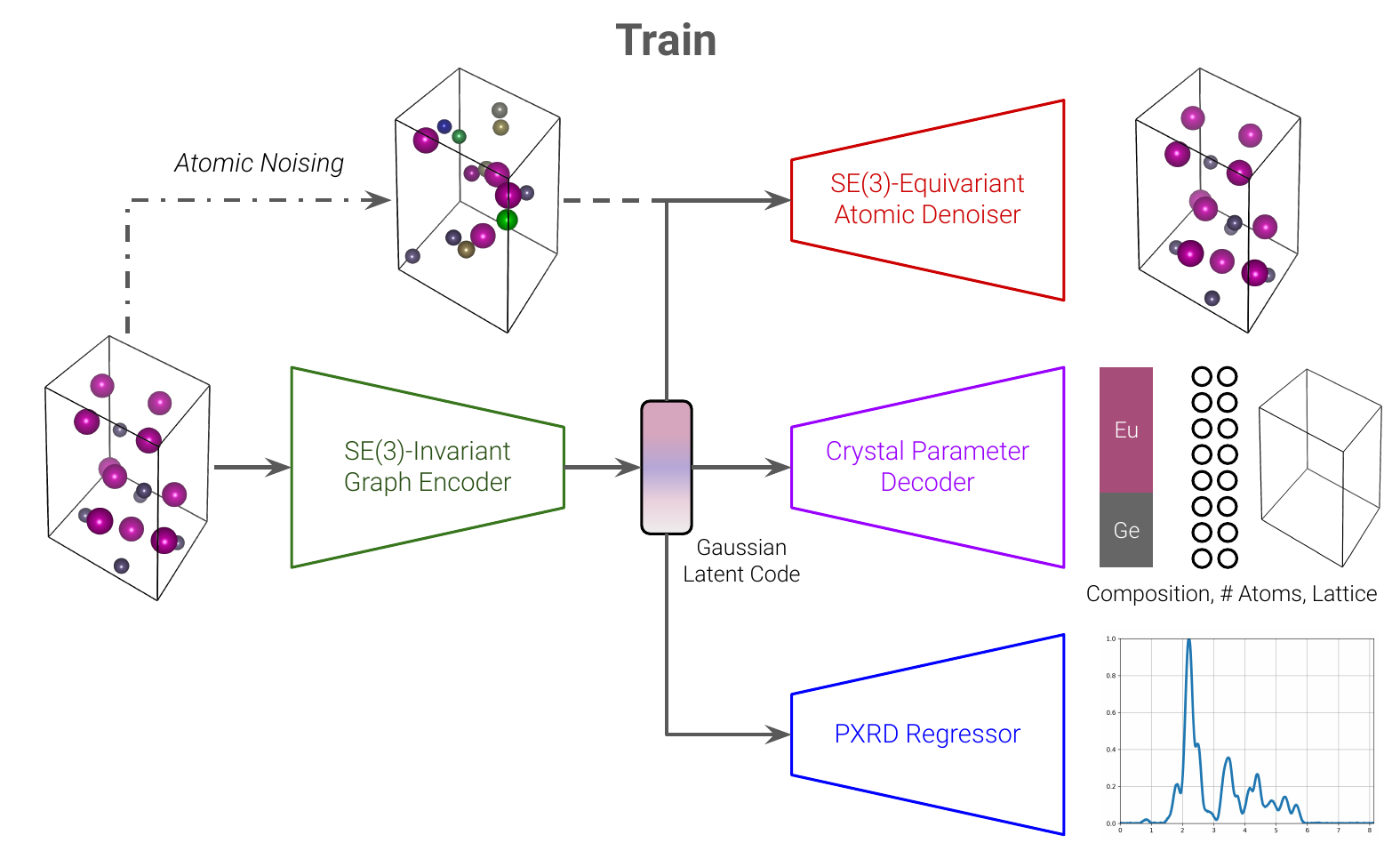}
    \caption{\textbf{\nanogen's Training Process:} \nanogen's training process essentially follows that from CDVAE. The objective is twofold: (1) to learn how to reconstruct the material after a noising process; (2) predict PXRD patterns from a compressed material representation. To this end, it has three branches: denoising diffusion of atomic species and coordinates (via "SE(3)-Equivariant Atomic Denoiser"); variational autoencoding of composition, number of atoms, and lattice parameters (via "Crystal Parameter Decoder"); and PXRD regression (via "PXRD Regressor"). These branches are all connected via the common Gaussian latent code (generated from "SE(3)-Invariant Graph Encoder").}\label{fig:train_process}
\end{figure}

\nanogen builds off of CDVAE (Crystal Diffusion Variational Autoencoder), a previously proposed material structure generative model \cite{cdvae_xie2021crystal}. The CDVAE model unconditionally generates materials represented by a graph encoding of the following tuple of characteristics: a list of atom types $\mathbf{A} \in \mathbb{R}^N$, a list of atom coordinates $\mathbf{X} \in \mathbb{R}^{Nx3}$, and a periodic lattice $\mathbf{L} \in \mathbb{R}^{3x3}$ where $N$ is the number of atoms in the unit cell. Concretely, the model decodes material composition and structure from randomly sampled noise via a diffusion process. By training on a diverse set of observed material structures, CDVAE learns to capture the underlying distribution of physically stable materials.

At a high level, the idea behind CDVAE is adopted from the variational autoencoder \cite{kingma2022autoencoding} and denoising diffusion networks \cite{yang2023diffusion}, which are generative models that learn to decompress data from noise. Essentially, such models are trained to encode data $x$ to a latent representation~$z$, then map it back to the original space to produce a reconstructed version of the original data sample $x$. The latent representation~$z$ is regularized such that the distribution $p(z)$ looks roughly like a standard normal distribution over the latent space. At inference time, generating samples simply involves sampling from $N(0,I)$ and decoding to the original data domain.

Implementationally, CDVAE is a hybrid between the noise-conditioned score network (the diffusion model) and the variational autoencoder (VAE). To understand the breakdown of the VAE and diffusion components, recognize that the unit cell of a material can be represented by four components: (1) chemical composition, (2) number of atoms, (3) lattice parameters, (4) atomic coordinates. 
CDVAE's first branch addresses components (1)-(3) with a variational autoencoder. 
The encoder is DimeNet \cite{dimenet_gasteiger2020directional}, an SE(3)-invariant graph neural network that maps the graph representation of a material to a latent representation~$z$. 
The graph representation has been modified to be a directed multi-graph that respects the inherent periodicity of materials \cite{xie2018crystal}. The latent representation,~$z$, is regularized to be a multivariate Gaussian, via the Kullback–Leibler-divergence loss \cite{kingma2022autoencoding}.
Then, from~$z$, it decodes (1) the chemical composition (as ratios), (2) the number of atoms, and (3) the lattice parameters ($a$, $b$, $c$, $\alpha$, $\beta$, $\gamma$). 
Each prediction is produced by a separate crystal parameter multi-layer perceptron 
that takes in the latent encoding~$z$. This latent code~$z$ is especially important, as it is later used as the material representation in all the other branches of the model, as described below. 

CDVAE's second branch addresses component (4) with denoising diffusion via noise conditioned score networks \cite{song2019generative, song2020generative}. It assumes that components (2)-(3) -- the number of atoms and lattice parameters -- are fixed. The forward process perturbs the atomic coordinates and atomic species with multivariate Gaussian noise. The reverse process is parameterized by GemNet \cite{gasteiger2021gemnet}, an SE(3)-equivariant graph neural network. It is conditioned on the latent code,~$z$, from the aforementioned material graph encoding; thus, it is crucial to learn an informative~$z$ 
in the VAE branch in order for this diffusion branch to work. The reverse process essentially predicts how to denoise the perturbed atomic coordinates and species via annealed Langevin dynamics, such that they move to their true positions and revert to their true species \cite{song2019generative}. Again, the output graph representation is a directed multi-graph that is compatible with the periodic nature of materials \cite{xie2018crystal}.

Then, at generation time, CDVAE generates materials as follows: First, sample a latent code $z \sim N(0, I)$ from the multivariate Gaussian distribution. This is decoded with the aforementioned crystal parameter MLPs to gain components (1)-(3): the lattice parameters, chemical composition, and number of atoms. This can be used to initialize a unit cell, where the atomic positions are randomly chosen, again from $N(0, I)$. Then, the atomic positions and species are refined through the Langevin dynamics SE(3)-equivariant graph denoising process, which is also conditioned on the latent code~$z$. Note that the lattice parameters and number of atoms stay fixed throughout the denoising process. This yields the generated material.

A natural question that arises is, why is it necessary to have both VAE and denoising diffusion components?
The rationale for using denoising diffusion is that, when predicting atomic coordinates, it actually has a theoretical connection to harmonic forcefields \cite{cdvae_xie2021crystal}. Furthermore, denoising diffusion has achieved high performance in other generative tasks \cite{song2019generative, song2020generative}.
However, the VAE portion is still necessary, because the traditional formulation of denoising diffusion \cite{song2019generative, song2020generative, yang2023diffusion} requires a fixed data representation dimensionality. In this case, the dimensionality of the data representation is (some multiple of) the number of atoms, so that cannot be denoised. To get around this, the VAE branch predicts the number of atoms, which is then used to initialize the denoising diffusion. 
% Why other parts?

For additional details, including training loss functions, input and output representations, and hyperparameters, we refer the reader to the original CDVAE paper: we adopt most of their settings, unless specified otherwise \cite{cdvae_xie2021crystal}. Also, refer to Figure \ref{fig:train_process} for a schematic diagram of the training process for our CDVAE-inspired model.

\subsection*{Property Optimization}

As an additional feature, CDVAE includes property optimization, which allows one to steer the generative process towards materials with desired properties. A differentiable property predictor (implemented as a neural network) is trained to regress the desired property from the latent molecule representation~$z$. To bias generation towards molecules with the desired property at inference time, gradient descent is used to adjust~$z$ (while the property predictor is frozen) to achieve the desired predicted property value. The optimized~$z$ is then used to initialize the material structure, and is further passed as conditioning to the atomic denoising process, as previously described. 

\subsubsection*{PXRD Regressor Architecture}
\begin{figure}
    \centering
    \includegraphics[width=0.9\textwidth]{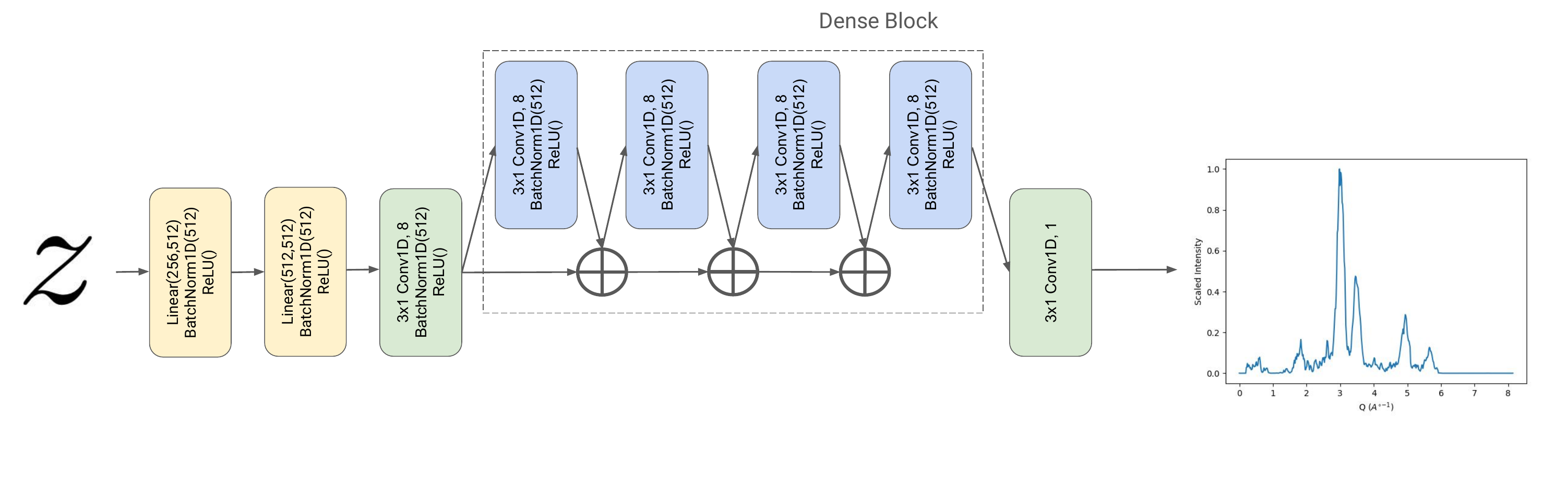}  
    \caption{\textbf{Visualization of \nanogen's PXRD Regressor:} Note $\oplus$ represents concatenation along the channel dimension. $F_{\psi}$ first projects latent material representations $z \in \mathbb{R}^{256}$ to a 512 dimensional representation via a sequence of fully connected linear layers, 1-dimensional Batch Normalization \cite{ioffe2015batch} and ReLU nonlinearity. A single 1-dimensional convolutional layer with a kernel size of 3, followed by a 1-dimensional Batch Normalization, is applied to the projection to produce an 8-channel representation. This representation kicks off the DenseNet feature aggregation, and is passed to a dense block consisting of 4 layers. Each layer maps aggregated features from previous layers in the block to a new 8-channel representation via 1-dimensional convolution with kernel size 3, followed by 1-dimensional Batch Normalization. The dense block produces an 8-channel, 512 dimensional signal. This gets passed through a final 1-dimensional convolutional layer with kernel size 3 to shrink the number of channels to 1 and produce the predicted $\textit{Q}$-space PXRD representation.}
    \label{fig: PXRD Regressor}
\end{figure}

In our setting, the PXRD pattern is the desired property to predict.
Thus, we design a PXRD Regressor $F_{\psi}$ that maps from \nanogen's latent material representations $z \in \mathbb{R}^{256}$ to a vector $y \in \mathbb{R}^{512}$, the estimated $Q$-space representation of the material's PXRD pattern. The PXRD Regressor is parameterized by a DenseNet-inspired architecture \cite{huang2018densely}, which extends traditional convolutional neural networks. We base our regressor off of the design in \textit{CrystalNet} \cite{crystalnet_guo2023towards}, which has a densely connected architecture with 1D inputs and outputs. Concretely, for a given depth in the network DenseNet aggregates previous intermediate data representations as input to the next convolutional layer. DenseNet has been shown to reduce the vanishing gradients problem and achieves strong results on standard computer vision benchmarks \cite{huang2018densely}. A visualization of \nanogen's PXRD Regressor is shown in Figure \ref{fig: PXRD Regressor}.

\subsection*{Generative Structure Solution Process}

Now, we describe our method for using the CDVAE-inspired model to generate structure solutions, and where we diverge from the original CDVAE formulation. Refer to Figure \ref{fig:test_process} for an overview.

\subsubsection*{Latent Code Selection via PXRD and Formula Guided Backpropagation}

\begin{figure}
    \centering
    \includegraphics[width=\textwidth]{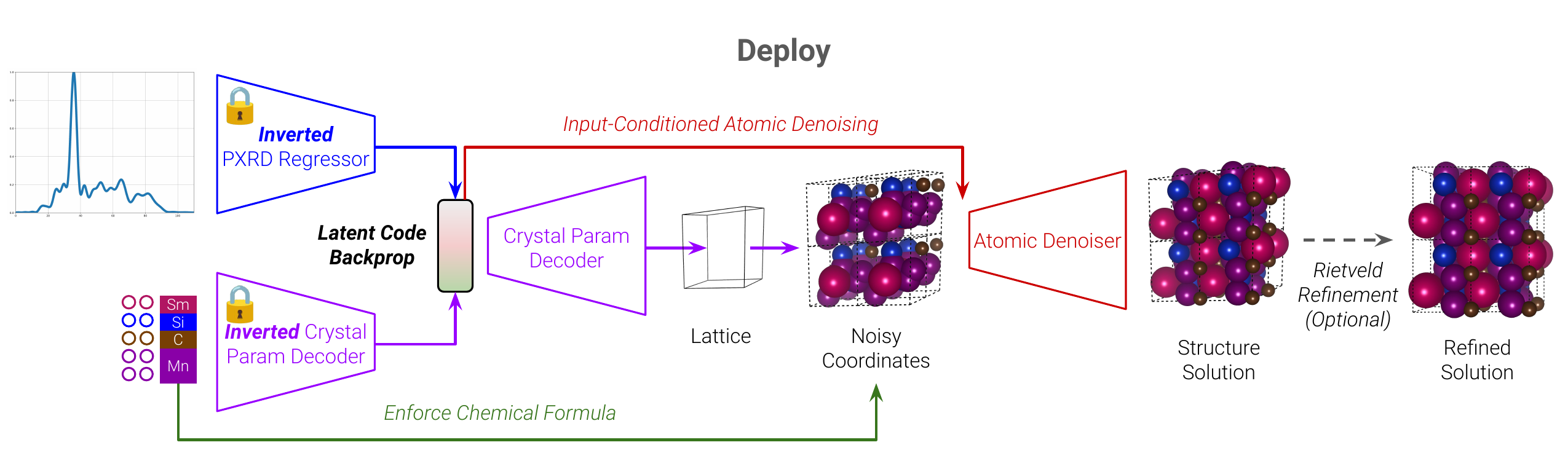}  
    \caption{\textbf{\nanogen's PXRD- and Formula-Guided Structure Solution Process:} \nanogen's material generation process uses the same components as the training process (Figure \ref{fig:train_process}), but in a different way. It inverts the PXRD Regressor and Crystal Parameter Decoder; in this way, conditioned on the PXRD and chemical formula, we can generate a latent code that fulfills these properties. We then use the latent code in the same way as the training process, feeding it through the Lattice Decoder and the Atomic Denoiser to generate the structure solution.} \label{fig:test_process}
\end{figure}

\begin{figure}
    \centering
    \includegraphics[width=\textwidth]{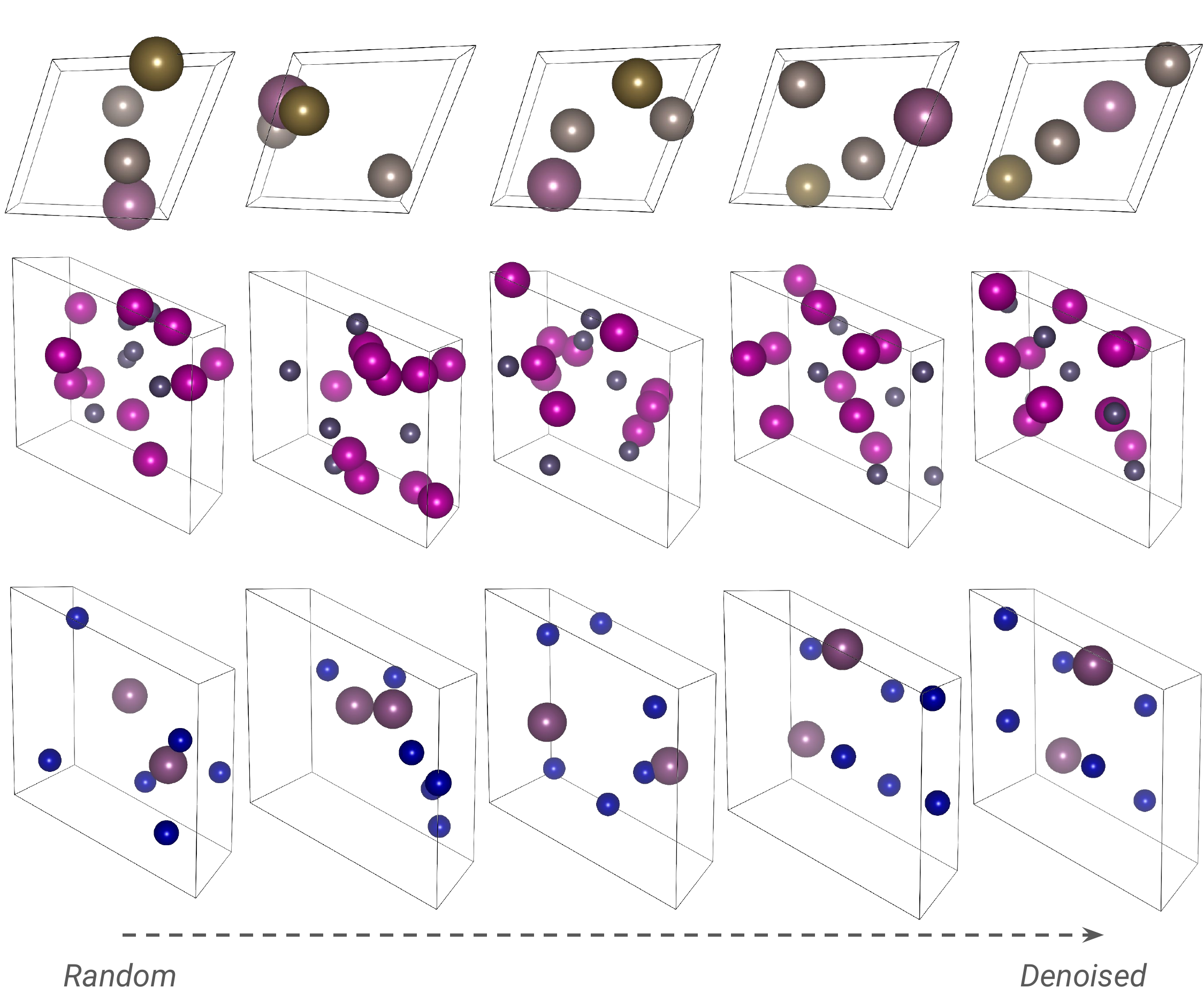}  
    \caption{\textbf{Visualization of Denoising Diffusion via Langevin Dynamics:} The leftmost column shows randomly initialized atomic coordinates within the unit cell. As we move right, we see the atomic coordinates gradually being denoised into a structural candidate.} \label{fig:diffusion_process}
\end{figure}

\nanogen modifies CDVAE by incorporating PXRD observations and chemical formulas as conditioning information for the generative process. This is achieved through the property optimization mechanism. 

For property optimization to work, at train time, we must first train models to predict the attributes we desire from a material, given its latent code. To this end we the PXRD regressor,~$F_\psi$, parameterized by a modified DenseNet architecture\cite{huang2018densely}. This is jointly trained with the base CDVAE model to predict the observed PXRD pattern of a material from its latent encoding~$z$. 
We also train MLPs, $F_\text{N}$ to predict the number of atoms in the unit cell, and $F_\text{comp}$ to predict the chemical composition from~$z$, as previously described in the variational decoding process. 

Then at test time we invert these models such that, conditioned on the attributes of interest (\textit{i.e.}, PXRD pattern, chemical formula), we can get latent codes that adhere to these attributes. 
Specifically, to generate candidate structure solutions given a PXRD pattern and chemical composition, we apply property optimization to a batch of randomly sampled latent vectors $z \sim N(0, I)$. Concretely, we freeze all parameters of \nanogen and apply gradient descent to adjust each~$z$ to optimize the objective,
\begin{equation}\label{eqn:loss}
    \min_{z} \quad \mathcal{L}_{\text{XRD}}(z, y, N, c) = \|F_{\psi}(z) - y\|_1 - \beta \log p(z) + \beta_{N}\mathcal{L}_{\text{softmax}}(F_{\text{N}}(z), N) + \beta_{C}\mathcal{L}_{\text{CE}}(F_{\text{comp}}(z), c),
\end{equation}
where $y$ is the PXRD pattern of the desired material, $N$ is the number of atoms in the unit cell, $c \in \mathbb{R}^{|A|}$ is normalized chemical composition, and $F_N$, $F_{\text{comp}}$ are the atom cardinality and composition prediction MLPs. The first two terms in $\mathcal{L}_{\text{XRD}}$ encourage a latent representation that corresponds to a material with the desired PXRD pattern $y \in \mathbb{R}^{512}$, while adhering reasonably to the Gaussian prior on the latent representations $p(z)$. The $\beta$ term controls the degree of adherence to the prior. The last two terms ensure that the latent representation corresponds to a material with the desired number of atoms and chemical composition, and $\beta_{N}, \beta_{C}$ control their respective influences.

We sample and optimize $100$ latent vectors~$z$. We use learning rate ranging from $0.1 \rightarrow 10^{-4}$ in a cosine decay with warm restarts setting (where each decay period length is doubled upon restart) \cite{loshchilov2016sgdr}, $\beta=2 * 10^{-4}$, $35000 = 5000 + 10000 + 20000$ total epochs, and $\beta_{N} = \beta_{C} = 0.1$.

We emphasize that the PXRD Regressor and Inverted Crystal Parameter Decoder in Figure \ref{fig:test_process} actually activate the same latent space encoding $z$. They can do this because of the combined loss in Equation \ref{eqn:loss}: the latent code is passed through both forward models, but rather than updating the model parameters in the gradient descent procedure, we update the latent code values based on the loss. In this sense, inversion of the models is not in the feed-forward sense, it is in the sense that the loss function is used to update the latent code parameters, rather than the model parameters as would be traditionally done.

\subsubsection*{Candidate Generation with Teacher Forcing}

We map the set of optimized latent representations~$z$ (in our setting, we use $100$) to materials with the decoding process described previously.~$z$ is used to initialize the lattice, and is used as conditioning for the atomic coordinate denoising process. 
However, we make a change from the originally described material generation process. We already know the chemical formula. So, we fix the number of atoms and atomic species according to the chemical formula, forgoing the need to use their respective MLPs to predict them again. 
This technique can be seen as teacher forcing with the strict chemical formula \cite{goodfellow2016deep, teacher_forcing_williams1989learning}, where the strict chemical formula has the absolute quantities of atoms in the unit cell, in addition to the ratios.

Further comparing the training process (Figure \ref{fig:train_process}) to the generative structure solution process (Figure \ref{fig:test_process}), we see that with the exception of the graph encoder, all the same components (atomic denoiser, crystal parameter decoder: composition + atomic cardinality + lattice, PXRD regressor, Gaussian latent code) are used. The difference is that in the structure solution process, the PXRD regressor, the composition decoder MLP, and the atommic cardinality decoder MLP are inverted. In other words, the input is now the PXRD pattern and chemical formula, and the output is a latent code that is optimized via backpropagation while these models are frozen. This results in a latent code that achieves high adherence to the inputted PXRD pattern and chemical formula. Then, the lattice parameter decoder and atomic denoiser are given this latent code (which should contain information about the material structure, based on the PXRD and chemical formula) to generate the lattice parameters and atomic coordinates as before.

\subsubsection*{Candidate Filtering via PXRD Simulation}

We then simulate the PXRD pattern for each of the $100$ candidates with the PyMatGen package \cite{pymatgen_ong2013python}. We simulate nanoscale shrinkage and apply Gaussian smoothing on these candidates' simulated PXRD patterns, just as was done for the ground truth PXRD. Finally, the top~$10$ candidates are selected based on the adherence (measured as $l_1$ loss) of simulated PXRD to the original, desired PXRD $y$. These~$10$ candidates are treated as \nanogen's estimate of possible structure solutions of the molecule given the PXRD pattern and chemical composition.
(In principle, we can have an arbitrarily high number of candidates, if we wish to expand our search.)

\subsection*{Training Loop}

Unless otherwise specified, we follow the hyperparameter settings in CDVAE \cite{cdvae_xie2021crystal}. The only major change we make is to adopt a cosine decay with warm restart learning rate schedule, where we have $1000$ total epochs, the learning rate ranges from $5 * 10^{-4} \rightarrow 10^{-6}$, the initial decay period is $65$ epochs, and the number of epochs per decay period is doubled on each restart \cite{loshchilov2016sgdr}.

\subsection*{Baseline: CDVAE Latent Space Search}
To our knowledge, there exist no \textit{ab initio} end-to-end structure solution methods for nanomaterials that we could use as benchmarks (\textit{i.e.}, PXRD pattern and formula as input, structure as output). The existing structure solution works are: (a) only for \textit{non}-nanosized materials and (b) do \textit{not} publicly release their code \cite{li2024powder, lai2024end, riesel2024crystal}. Thus, to assess the ability of \nanogen, we conduct an ablation on the main modification we made to the CDVAE model: the ability to condition on PXRD patterns.

In particular, we want to investigate how beneficial our latent code selection process is at creating materials that adhere to the given PXRD and formula information. Thus, we keep all the same parts of our generative structure solution process, except that we do \textit{not} use PXRD or formula information to optimize the latent codes via backpropagation. 

Instead, we randomly sample the same number ($100$) of latent codes from the same distribution $N(0, I)$ as before. Then, we skip the latent code optimization, and go directly to the decoding process, which is conducted in exactly the same way as our method. That is, we use the same model with the same noise schedule, and we keep the teacher forcing, such that the atomic species and number of atoms are fixed to the true chemical formula. Finally, we do the same candidate filtering with PXRD simulation, to see which of the generated materials matches the ground truth PXRD pattern. Again, we keep the top~$10$ candidates, and calculate evaluation metrics in exactly the same way as for our method.

% \subsubsection*{Semi-Random Material Construction}

% For our next ablation, we want to investigate the importance of the atomic denoising process. So, in addition to removing latent code optimization, we also remove the atomic denoising process. 

% It is conducted as follows. We randomly sample $100$ latent vectors $z \sim N(0, I)$, the same number as in the regular material generation process. Without any optimization, we then use these latent vectors (via decoding from our trained MLP) to initialize the unit cells of the materials. As in the other evaluations, we assume that we know the chemical formula, and we randomly initialize the atomic positions, as in our method. The difference is that we do not conduct any denoising on these positions. Then, in exactly the same way as our method, we calculate the PXRD adherence between each of these $100$ materials and the ground truth PXRD, and choose the top~$5$ candidates. From there, evaluation metrics are calculated in the same way as previously described. 

% This is semi-random, because while~$z$ is randomly drawn from the latent distribution, once it is decoded by the MLP, it does contain priors about physically plausible unit cell shapes. For this reason, this is actually a generous baseline, because it makes use of a small portion of our trained model. Essentially, the only differences between this baseline and our method are (1) the lack of PXRD-guided latent code optimization; (2) the lack of denoising. All other parts are the same, making for a fair comparison.

\subsection*{Metric: R-Factor}

To quantify \nanogen performance, we use the R-Factor (residuals function), which is the standard metric used by crystallographers when determining the success of their structure solutions \cite{billinge2008powder_diffraction, egami2012underneath}. It operates directly on the reciprocal space diffraction patterns: the advantage of this is that when deployed in real-world experimental settings, it is directly observable. 

We calculate the R-factor (residuals function) defined as,
\begin{equation}
    R_{wp}^{2} = \frac{\int[I_\text{gt}(Q) - I_\text{pred}(Q)]^{2}dQ}{\int[I_\text{gt}(Q)]^2dQ},
\end{equation}
where $I_\text{gt}(Q)$ is the ground truth PXRD pattern in $Q$-space, and $I_\text{pred}(Q)$ is the PXRD pattern in $Q$-space simulated from the predicted crystalline structure. (Sometimes, peaks corresponding to certain Q-values are up-weighted in the calculation \cite{egami2012underneath}; but for simplicity, we give every point equal weight.) The PXRD patterns are computed on a grid of $8192$ evenly spaced samples in $Q$ space.
Typically, \rw below $10\%$ can be considered to be a success \cite{billinge2008powder_diffraction}. 

When we calculate this metric, we calculate the PXRD patterns \textit{without} any simulated nanoscale shrinkage. The reason why we omit this is because when comparing the 10~\AA\ experiments to the 100~\AA\ experiments, the different sinc filters would result in different peak broadening widths, and give a non-consistent standard of comparison across these experiments.  

Instead, we consistently apply a Gaussian filtering on the PXRD patterns with $\sigma = 0.040784 \text{~\AA}^{-1}$ (this is only $0.5\%$ of the PXRD pattern width). The rationale for this is that if we were to use the raw PXRD patterns without any peak broadening, a negligible difference in peak location (let us say, $10^{-6}$~\AA) would actually lead to a very poor \rw, since the integral in the numerator calculates the error at each infinitesimally small location. However, it should be clear that such an example is indeed a good fit. So, to avoid this problem, we add the slight Gaussian peak broadening, such that if the peaks are "close enough", we will still have low error values.
We note that varying the width of this Gaussian will change the resulting agreement factors, and so the absolute value of this measure does not carry definitive information, but the relative value between solutions certainly does. 

We note that this way of computing the R-factor is an unusually strict metric since we score model success by comparing to data of higher resolution than was actually shown to the model.

Finally, we report the best \rw among the~$10$ \nanogen candidates for each material. This is reasonable, because as stated earlier, the ground truth PXRD can be directly observed in experimental settings.

%\subsubsection*{Real-World Evaluation}

\subsection*{Structural Refinement Pipeline}

As described in the text we carry out a Rietveld refinement on the results of a small subset of materials as would be done in a conventional structure solution procedure.

\subsubsection*{Material Curation}

We uniformly selected $20$ materials for Rietveld refinement from the test set. For each of these materials, we took the ten best candidates generated by \nanogen, and refined all of them. 

\subsubsection*{PXRD Pre-Processing}

We used the PXRD patterns from both the simulated 10~\AA\ and 100~\AA\ nanomaterial sizes. We convolved the 10~\AA\ patterns with a Gaussian filter with low sigma (0.081568~\AA$^{-1} = 1$\% of pattern length in $Q$), and we convolved the 100~\AA\ patterns with a lower-sigma filter (0.040784~\AA$^{-1} = 0.5\%$ of pattern length in $Q$). 

\subsubsection*{Rietveld Refinement Parameters}

Since the lattice parameters and space group symmetry (all in \textit{P1}) of the predictions are different than the ground truth, the simulated diffraction patterns of the predictions are typically different than those of the ground truth despite the topological similarity. 
We performed a constrained search over distortions of the predicted unit cell via refinements using the program TOPAS v7 \cite{coelho2018topas}. 
Due to the different simulation engine, we first ensured a suitable profile to describe the peak-shape by fitting the ground truth model to the ground truth simulated data using a Gaussian convolution for the 100~\AA\ and the 10~\AA\ patterns. 
The resulting values were then fixed for refinements of \nanogen candidates, which had been simulated with the same Gaussian filter as the ground-truth patterns. 
For each predicted structure, all lattice parameters were then allowed to refine within a window of +/- 20\%\, along with a scale factor, where the parameters were randomized after each convergence to escape local minima over 2000 iterations. 
Due to the low model symmetry and low angular range and high broadening of the patterns, we did not attempt to additionally refine the atom site positions. 
This is in contrast to the final refinement step in conventional structure solution from crystalline materials and so we do not expect to obtain \rw %\sjb{check we use rw consistently everywhere, not sometimes squared and sometimes not} 
values as good as those from full refinements, even for correct structures. (We note that TOPAS uses a slightly different calculation of \rw that incorporates statistical weighting, but nonetheless, good results with the TOPAS calculation correspond to good results with our calculation.)
% The $R_{wp}$ value was calculated as given in TOPAS by
% \begin{equation}
%     R_{wp} = \sqrt{\sum{w_m \left(Y_{o,m} - Y_{c,m}\right)^2/\sum{w_m Y_{o,m}^2}}}
% \end{equation}
% where $Y_{o,m}$ are the target intensities, $Y_{c,m}$ are the calculated intensities, and $w_m$ is the statistical weighting. 

\subsubsection*{R-Factor Comparison}

Finally, we compare the pre-refinement \rw to the post-refinement \rw. A goal here is to look for correlations between metrics for structures that are successful pre and post refinement.  
If there is a strong correlation we can use pre-refinement metrics as a low cost proxy upper bound for models that will perform well post-refinement.

Although during the Rietveld refinement, we optimize \rw from the material degraded by concurrent sinc and Gaussian convolutions, for this comparison, we compare all the \rw (both pre- and post- refinement) \textit{without} any simulated nanoscale broadening but with the low-sigma Gaussian broadening only. We believe that this gives a better assessment of the true quality of the structure solution. %See "Metrics/R-Factor" for more details on the calculations.

\bibliography{sample}

\section*{Acknowledgements}
%Funding.
We thank Judah Goldfeder for helpful discussions regarding the viability of a diffusion-based approach. We thank Ethan Bondick and Hao Lin for their assistance in preliminary materials visualization.

\section*{Author contributions statement}

SJLB and HL proposed the research topic. GG and TLS proposed the machine learning methodology and wrote the code. GG, TLS, MV, and SJLB conducted the nanomaterial shrinkage analysis. GG and SJLB procured the data. GG, SJLB, and HL designed the experiments. MWT conducted structural refinement. All authors wrote the manuscript and provided feedback.

\section*{Additional information}

Work in the Lipson group was supported by U.S. National Science Foundation under AI Institute for Dynamical Systems grant 2112085. Work in the Billinge group was supported by the U.S. Department of Energy, Office of Science, Office of Basic Energy Sciences (DOE-BES) under contract No. DE-SC0024141. 
This material is based upon work supported by the U.S. Department of
Energy, Office of Science, Office of Advanced Scientific Computing Research, Department of
Energy Computational Science Graduate Fellowship under Award Number DE-SC0025528 to G. Guo.

\subsection*{Code Availability}

Please obtain the code, with reproducibility instructions, at: \href{https://github.com/gabeguo/cdvae_xrd}{https://github.com/gabeguo/cdvae\_xrd}.

\subsection*{Data Availability}

MP-20-PXRD is available at: \href{https://github.com/gabeguo/cdvae_xrd/tree/main/data/mp_20}{https://github.com/gabeguo/cdvae\_xrd/tree/main/data/mp\_20}. Instructions for obtaining the experimental data are available at: \href{https://github.com/gabeguo/cdvae_xrd/tree/main/data/experimental_xrd}{https://github.com/gabeguo/cdvae\_xrd/tree/main/data/experimental\_xrd}.

\end{document}